\documentclass[ trackchanges,twocolumn]{aastex701}

\usepackage{amsmath}
\usepackage[export]{adjustbox}
\usepackage{array}
\usepackage{hyperref}
\usepackage{comment}

\received{May, 2026}
\accepted{July, 2026}
\submitjournal{The Astrophysical Journal}

\begin{document}

\title{A cooler look at the environment of Cygnus X-1: \\Searching for dynamical interactions within cold molecular gas}

\author[orcid=0000-0002-1514-5558,sname='Bosch-Cabot']{Pau Bosch-Cabot}
\affiliation{Department of Physics and Astronomy, University of Lethbridge, Lethbridge, Alberta, T1K 3M4, Canada}
\email[show]{pau.boschcabot@uleth.ca}  

\author[orcid=0000-0003-3906-4354,gname=Alexandra, sname='Tetarenko']{Alexandra J. Tetarenko} 
\affiliation{Department of Physics and Astronomy, University of Lethbridge, Lethbridge, Alberta, T1K 3M4, Canada}
\email{}

\author[orcid=0000-0002-6043-5079,gname=Valenti, sname='Bosch-Ramon']{Valentí Bosch-Ramon} 
\affiliation{Departament de Física Quàntica i Astrofísica, Institut de Ciències del Cosmos Universitat de Barcelona (ICCUB), Universitat de Barcelona (IEEC-UB), E08028 Barcelona, Catalonia, Spain}
\email{}

\author[orcid=0000-0003-3124-2814,gname=James CA, sname='Miller-Jones']{James C. A. Miller-Jones} 
\affiliation{International Centre for Radio Astronomy Research -- Curtin University, Perth, WA 6845, Australia}
\email{}

\author[orcid=0000-0002-3500-631X,gname='David', sname='Russell']{David Russell} 
\affiliation{Center for Astrophysics and Space Science, New York University Abu Dhabi, PO Box 129188, Abu Dhabi, UAE}
\email{}

\author[orcid=0009-0003-5079-5139,gname=Sara E., sname='Motta']{Sara E. Motta} 
\affiliation{Istituto Nazionale di Astrofisica, Osservatorio Astronomico di Brera, Via E. Bianchi 46, 23807 Merate (LC), Italy}
\affiliation{Università degli Studi di Milano Bicocca, Dipartimento di Fisica, Piazza dell’Ateneo Nuovo, 1 – 20126 Milano Casella, Italy}
\email{}

\author[orcid=0000-0001-8125-5619,gname=Pikky, sname='Atri']{Pikky Atri} 
\affiliation{ASTRON, Netherlands Institute for Radio Astronomy, Oude Hoogeveensedijk 4, 7991 PD Dwingeloo, The Netherlands}
\affiliation{Department of Astrophysics/IMAPP, Radboud University, P.O. Box 9010 6500 GL Nijmegen, The Netherlands}
\email{}

\author[orcid=0000-0001-7796-4279,gname=Maria, sname='Diaz-Trigo']{María Díaz-Trigo} 
\affiliation{ESO, Karl-Schwarzschild-Strasse 2, 85748 Garching bei München, Germany}
\email{}

\author[orcid=0009-0003-5079-5139,gname=Isabella, sname='Mariani']{Isabella Mariani} 
\affiliation{Istituto Nazionale di Astrofisica, Osservatorio Astronomico di Brera, Via E. Bianchi 46, 23807 Merate (LC), Italy}
\affiliation{Università degli Studi di Milano Bicocca, Dipartimento di Fisica, Piazza dell’Ateneo Nuovo, 1 – 20126 Milano Casella, Italy}
\email{}

\author[orcid=0000-0003-3165-6785,gname=Steve, sname='Prabu']{Steve Prabu} 
\affiliation{University of Oxford, Department of Physics, Astrophysics, Denys Wilkinson Building, Keble Road, OX1 3RH Oxford, United Kingdom}
\affiliation{International Centre for Radio Astronomy Research -- Curtin University, Perth, WA 6845, Australia}
\email{}

\begin{abstract}
 
We present IRAM--30m observations aimed at identifying potential outflow-interstellar medium interaction sites in the vicinity of the black hole X-ray binary Cygnus X--1, which displays persistent relativistic jets and a prominent stellar wind. Using this dataset, we construct molecular line emission maps, identifying a never before seen molecular structure potentially linked to X-ray binary-driven feedback. This structure, surrounding the source, exhibits properties consistent with an interaction powered primarily by the stellar wind of the massive stellar companion and further sculpted by the relativistic jets. Our results indicate that multiple outflow mechanisms (stellar winds and relativistic jets) may simultaneously be shaping the interstellar medium around Cygnus X--1, and that molecular line imaging can help to disentangle complex feedback processes in environments where multiple outflows take place. 

\end{abstract}



\section{Introduction} 

Relativistic jets launched from accreting black holes (BHs) deposit large amounts of energy and matter into their environment, and in turn can significantly influence the environment's evolution. Jet feedback comes in a variety of forms. In super-massive black holes at the centers of active galactic nuclei (AGN), jets are often observed to carve out huge cavities in hot gas, in turn potentially affecting the entire galaxy's evolution \citep[e.g.,][]{magorrian1998demography,mcnamara2007heating, nesvadba2010energetics, krause2023jet}. 
The stellar-mass analogues of AGN, black hole X-ray binaries (BHXBs), similarly have a significant effect on the surrounding interstellar medium (ISM). For example, BHXB jets are known to play a significant role filling the galactic disk and the halo of the Galaxy with magnetized plasma \citep{heinz2008blazing}, as well as ultimately accounting for the average magnetic field observed in the Galaxy \citep{alfaro2024ultra}. 

Despite the clear imprint of jets on their environments from stellar to galactic scales, the jet-launching mechanisms and power source remain one of the central unresolved problems in BH astrophysics. 
The physical mechanism responsible for launching BH jets has been thoroughly studied over the last few decades. Current models attribute their origin either to the extraction of rotational energy from a spinning BH \citep{blandford1977electromagnetic} or to magnetohydrodynamic processes tapping the angular momentum of the material accumulated around the BH in the accretion disk \citep{blandford1982hydromagnetic}. Discriminating between these possibilities hinges on reliable and independent estimates of jet power. Despite recent results with Very Long Baseline Interferometry \citep[VLBI;][]{prabu2026jet}, obtaining these estimates from direct observations usually remains quite challenging, given that the jets are radiatively inefficient and move at relativistic speeds. 

Nevertheless, a promising avenue to constrain BH jet energetics is the study of jet interactions with the surrounding environment. The characterization of the affected gas enables us to establish a connection between the local conditions of the ISM and the lifetime averaged output of jet power \citep{kaiser1997self, kaiser2004revision}. This technique (often called calorimetry) has been used on several AGN sources through studies of their extended radio lobes \citep[e.g.,][]{churazov2002cooling}. Additionally, BHXBs have proven to be ideal targets for studying the interactions between the jet and the ambient medium, due to their close proximity and the rapid timescale evolution of BHXB jet activity. 
The calorimetry approach has been conducted in a handful of BHXBs, relying on the study of continuum radio structures \citep[e.g.,][]{gallo2005dark, atri2025quantifying, motta2025meerkat, mariani2025meerkat}, optical line emission \citep[e.g.,][]{zealey1980interaction, russell2007jet, sell2015shell, mcleod2019optical}, mapping of molecular line emission \citep[i.e., astrochemistry, e.g.,][]{tetarenko2018mapping,tetarenko2020jet,bosch2026constraining}, and even gamma-ray detections \citep[e.g.,][]{hess2024acceleration}.

The majority of BHXB systems targeted with the calorimetry approach host BHs with low-mass stellar companions (low-mass X-ray binaries; LMXBs), in which the dominant source of long-term kinetic feedback into the ISM is the relativistic jet itself. During LMXB transient outbursts, mass-loaded accretion disk wind outflows can also be a relevant feedback source \citep[e.g.,][]{munoz2016regulation}. In contrast, BHXBs with high-mass stellar companions (high-mass X-ray binaries; HMXBs) exhibit a more complex long-term feedback scenario, as there exist persistent outflows from the stellar companion in the form of stellar winds \citep{puls2008mass}. These stellar winds are thought to produce comparable feedback effects to jets when interacting with the ISM \citep[e.g., shocked gas regions;][]{comeron1998numerical, bosch2011termination, martinez2023probing}. When the ISM is affected by multiple different outflows, the classical continuum imaging approach is not informative enough to disentangle the effects of each component on the ISM.

The observation of astrochemical tracers is particularly insightful for studying complex and entangled interactions. Molecular line emission can be used to identify where the gas mass is located in a region, the presence of shocks, and even reveal complex gas kinematics that can be traced back to outflow-ISM interactions through the spectral analysis of the emission \citep{tetarenko2018mapping,tetarenko2020jet,bosch2026constraining}. In particular, since jets and winds introduce feedback with different dynamics (e.g., collimated, relativistic jets vs. isotropic, non-relativistic winds), their effect on the surrounding environment could have distinct signatures detectable within molecular line properties which are not traceable with continuum images alone.
In this work, we study the molecular gas emission in the vicinity of the Cygnus X--1 BHXB system. 

Unlike the previous BHXBs whose environment has been observed with this astrochemical approach, Cygnus X--1 shows not only persistent relativistic jets, but also a powerful stellar wind from its massive companion star \citep[e.g.,][]{herrero1995fundamental, brigitte2025disentangling}. As a result, two distinct and persistent outflow mechanisms operate simultaneously, each capable of injecting large amounts of energy into the surrounding ISM \citep{sell2015shell}.

\subsection{Cygnus X--1}
Cygnus X--1 (hereafter Cyg X--1) is a HMXB located at a distance of $2.2\pm0.2$ kpc, and consisting of a $21.2\pm2.2~M_\odot$ BH and a spectral type O supergiant star of mass $41\pm8~M_\odot$ \citep{miller2021cygnus} in a 5.6 day orbit \citep{brocksopp1998improved}. The system has a well constrained 3-dimensional velocity, including measurements of the heliocentric radial velocity of the system ($-5.1\pm0.5~\text{km s}^{-1}$, \citealt{gies2008stellar}) based on optical spectroscopy, and an apparent proper motion of $\mu~(\text{RA,DEC})\simeq (-3.8,-6.3)~\text{mas yr}^{-1}$ \citep{miller2021cygnus} constrained by radio VLBI observations. The proper motion translates into an apparent movement in the south of west direction of the sky plane. This motion implies a peculiar velocity outside of galactic rotation of $v_{\rm pec}\sim20~\text{km s}^{-1}$ \citep{carretero2025observational}, suggested to be even lower when compared to the mean velocity of surrounding stars \citep{nagarajan2025mixed}.  

Cyg X--1 is particularly well studied with detailed X-ray monitoring, and is known to spend significant amounts of time in an accretion state with persistent, compact jets \citep[e.g,][]{fender2006transient, grinberg2013long}. In addition, the massive stellar companion displays a prominent stellar wind, which acts as the main source of material for accretion onto the BH. This stellar wind has a mass loss rate of $\dot{M}_w\sim10^{-6}~\text{M}_\odot~\text{yr}^{-1}$ and a terminal wind velocity of $v_w\sim1800-2600~\text{km s}^{-1}$ \citep{herrero1995fundamental,gies2003wind,lai2024characterisation,ramachandran2025comprehensive, brigitte2025disentangling}. Further, this stellar wind is known to interact with the jet, causing a bending effect, which when coupled with the orbital motion of the system results in a helical propagation pattern of the jet \citep{prabu2026jet}.

Radio continuum observations of the environment around Cyg X--1 have revealed emission from a bow shock structure, that has been attributed to the leading edge of a cocoon inflated by either the approaching component of the jet, the powerful stellar wind of the companion star, or a combination of the two \citep{marti1996search, gallo2005dark, sell2015shell}. This same region was found to be bright in optical line emission through observations targeting ionized gas \citep{russell2007jet}. A search for diffuse X-ray emission in this region showed no significant emission, and the combination of optical line and radio continuum observations yielded constraints on the jet power and the age of the structure through calorimetry ($\leq4\times10^4~\text{yr}$; \citealt{russell2007jet, sell2015shell}). 
The formation of a clear bow shock in the direction of only the approaching jet may be explained by the proximity to the Tulip nebula H{\sc ii} region (north east from the source), where the local ISM density increases \citep{gallo2005dark}. On the other hand, no bow shock has ever been found in the receding jet direction, presumably because of local inhomogeneity in the ISM density distribution \citep{marti2017galactic}. 

\begin{figure*}[t]
    \centering
    \includegraphics[width=0.48\textwidth]{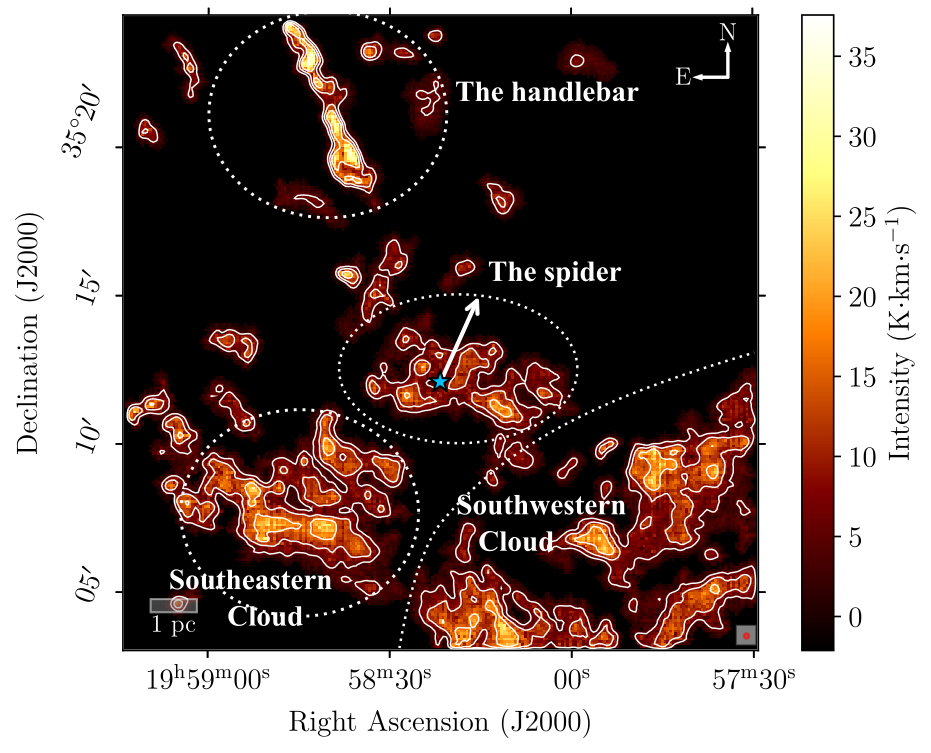}\hfill
    \includegraphics[width=0.48\textwidth]{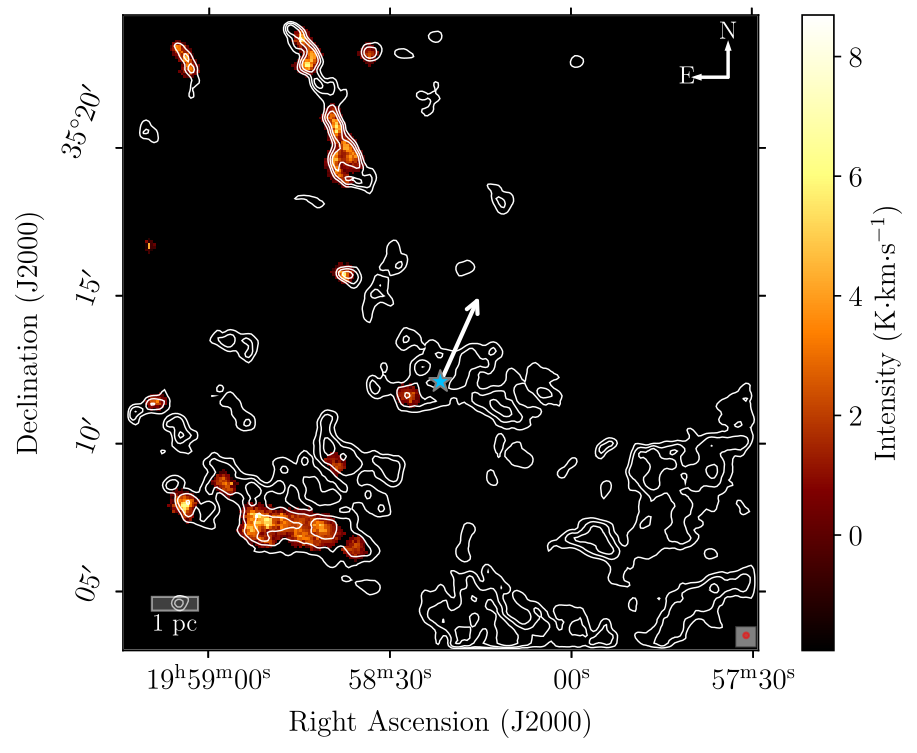}\hfill
    \caption{Integrated intensity maps of the $^{12}$CO (\textit{left}) and $^{13}$CO (\textit{right)} $J=2-1$ transition covering the Cyg X--1 region. The red ellipse at the lower right represents the IRAM--30m beam size, the position of the BHXB is represented as a blue star, and the approaching jet direction is represented as a white arrow. The contours trace the $^{12}$CO emission with levels $[4.8, 9.6, 16.3]$ K km s$^{-1}$. The main emission regions for this analysis are annotated in the \textit{left} panel. We identify several regions of bright molecular emission in this region, including a structure surrounding the central BHXB.}
    \label{fig:momentmaps}
\end{figure*}

There have been several more recent followup campaigns in the Cyg X--1 region. The Atacama Large Millimeter/Sub-millimetre Array (ALMA) surveyed a region at the edge of the bow shock in a search for SiO, H$_2$O, and CH$_3$OH molecular line emission associated to outflow based interactions. Non-detections of those tracers set solid upper limits to line emission around the edge of the bow shock \citep{trigo2021search}. Further, the bow shock has been revisited in the radio band leveraging MeerKAT's superior surface brightness sensitivity, providing observations that were more sensitive to faint structures and provided a wider frequency lever arm than past campaigns. In that work, the bow shock was detected in two radio bands, allowing the mapping of the spectral index, which suggested an inhomogeneity in the emission at the jet impact site that was previously unknown \citep{atri2025quantifying}.

In this paper, we report on observations of CO molecular line emission with the Institut de Radioastronomie Millim\'etrique 30m telescope (IRAM--30m), to characterize the physical and chemical properties of the ISM in the vicinity of Cyg X--1, and to search for signatures of dynamical interactions attributable to the stellar wind and the relativistic jets. Our primary goal is to assess whether the large-scale influence of these two outflows can be disentangled observationally, and to evaluate the extent to which astrochemical diagnostics can inform calorimetric estimates in such a complex feedback environment.
This paper is organized as follows. In \S\ref{sec:obs}, we describe the observations and data reduction procedures. In \S\ref{sec:res}, we present the results of our analysis. We discuss the implications of our findings and outline prospects for future work in \S\ref{sec:disc}.

\begin{figure*}[t]
    \centering
    \includegraphics[width=0.48\textwidth]{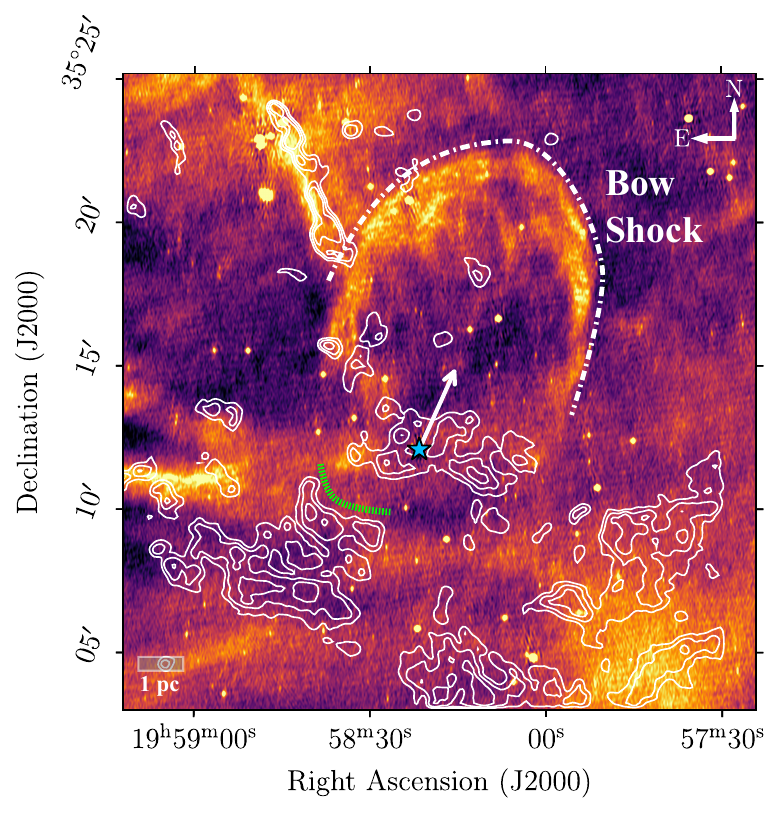}\hfill
    \includegraphics[width=0.48\textwidth]{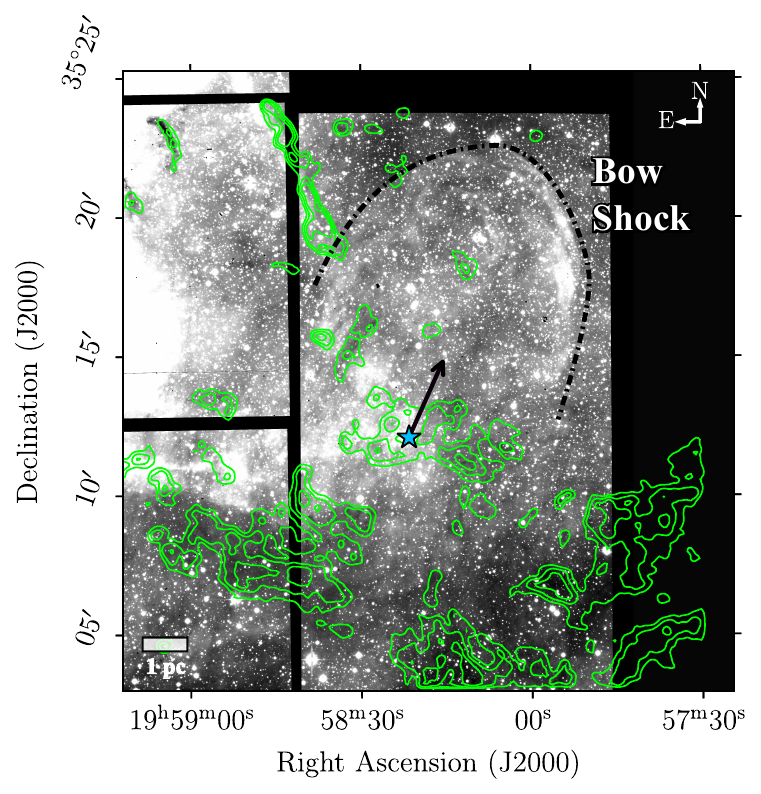}\hfill
    \caption{Images of the Cyg X--1 region taken with the MeerKAT radio telescope (S-band, 2.625 GHz central frequency; \textit{left}) and Isaac Newton Telescope (H$\alpha$; \textit{right)}. The position of the central BHXB is indicated with a blue star and the approaching jet direction is indicated with a white/black arrow. White and green contours trace the $^{12}$CO emission levels at $[4.8, 9.6, 16.3]$ K km s$^{-1}$. On the \textit{left} panel we also indicate an indentation shape of the {\it Southeastern Cloud} --reminiscent of a bow shock-- with a vertically-dashed green line. The region around and inside the radio/optical-bright bow shock structure (white/black dash-dotted line) is devoid of molecular gas within the probed velocity range, whereas the surroundings of the central BHXB source and the southern region show extended molecular features. }
    \label{fig:comparison}
\end{figure*}

\section{Observations and Data Analysis} \label{sec:obs}
\subsection{IRAM--30m observations}
Molecular gas observations were conducted with the IRAM--30m telescope (Sierra Nevada, Spain) on 2018 December 12 using the Heterodyne Receiver Array (HERA) multi-pixel receiver. We targeted the $^{12}$CO($J=2-1$) and $^{13}$CO($J=2-1$) transitions, observing at central rest frequencies of 230.538 GHz and 220.399 GHz, respectively, corresponding to an angular resolution of approximately $11''$. These transitions are known density tracers, and their relative intensity also traces optical thickness. Calibration and data reduction were performed using the \textsc{gildas} software\footnote{Software developed and maintained by IRAM: \hyperlink{}{https://www.iram.fr/IRAMFR/GILDAS/}} \citep{gildas}, applying the standard IRAM chopper-wheel method to yield spectra in the antenna temperature scale ($T_\mathrm{A}^\ast$). The resulting data were further analyzed by following standard procedures including baseline removal, flagging of bad channels, spectral regriding, and the construction of data cubes. Intensities were converted to main-beam brightness temperature using the relation $T_{MB}=\frac{F_{eff}}{B_{eff}}T_A^*$, where $F_{eff}=0.92$ and $B_{eff}=0.58$ at 230 GHz. 
Lastly, we applied a signal mask to all of the images presented in this work, which suppresses the contributions of noise\footnote{The sigma values are different for $^{12}$CO ($\sim0.7~\text{K/pixel/channel}$) and the $^{13}$CO ($\sim0.4~\text{K/pixel/channel}$).}. For this mask, we use the imaging strategy described in \cite{leroy2021phangs}, which creates a mask requiring the detection of a signal in two adjacent velocity channels at $\geq3\sigma$.

\subsection{Velocity range and kinematic distance considerations}

All radial velocities in this work are given with respect to the local standard of rest (LSR), with negative and positive values corresponding to motion toward and away from the reference frame, respectively. The spectral cubes produced from the IRAM--30m data explore the range $v_{\text{LSR}}=[-100,+100]~\text{km s}^{-1}$. We note that previous estimates of the bulk velocity of the bow shock's leading edge place it at $v\sim190-270~\text{km s}^{-1}$ \citep{sell2015shell}. While the projected radial velocity of the bow shock would exceed our observed velocity range, previous analyses of molecular emission near jet-ISM interaction regions have repeatedly found much slower components of molecular gas accumulating at the edges of expanding cavities \citep[e.g.,][]{tetarenko2018mapping,tetarenko2020jet}. Even if part of the emission might be missed, should there be any molecular gas clouds at the edge of the cavity reaching the bulk velocity  of the bow shock, projection effects may cause some extended emission regions to be observable within our probed velocity range. 

The potential association of any observed region to a BHXB outflow-ISM interaction relies on the proximity of the BHXB to the suspected interaction region. To assess this, we applied the kinematic distance estimation methods outlined in \cite{wenger2018kinematic}, utilizing the Galactic rotation parameters from \cite{reid2019trigonometric}.
Constraints from VLBI measurements \citep{miller2021cygnus} locate Cyg X--1 at $D=2.22^{+0.18}_{-0.17}$ kpc. In Galactic rotation and Cyg X--1's direction this corresponds to a radial velocity with respect to the LSR between 15 km s$^{-1}$ and 17 km s$^{-1}$. However, Cyg X-1 has a known heliocentric radial velocity of $\sim-5~\text{km s}^{-1}$ \citep{gies2008stellar}, corresponding to $v_{\text{LSR, Cyg X-1}}\simeq10.2\pm0.5~\text{km s}^{-1}$\citep{schonrich2010local}. In turn, this creates a range of radial velocities close to $v_{\text{LSR, Cyg X-1}}$ that may include gas co-moving or affected by the source. In our spectral cubes, we only find significant ($3\sigma$ ) emission in the [0,20] km s$^{-1}$ $v_{\rm LSR}$ range, and therefore we analyze all of this detected emission under the assumption that it could potentially have been affected by some form of the BHXB's feedback.

\begin{figure*}[t]
    \centering
    \includegraphics[width=0.60\textwidth]{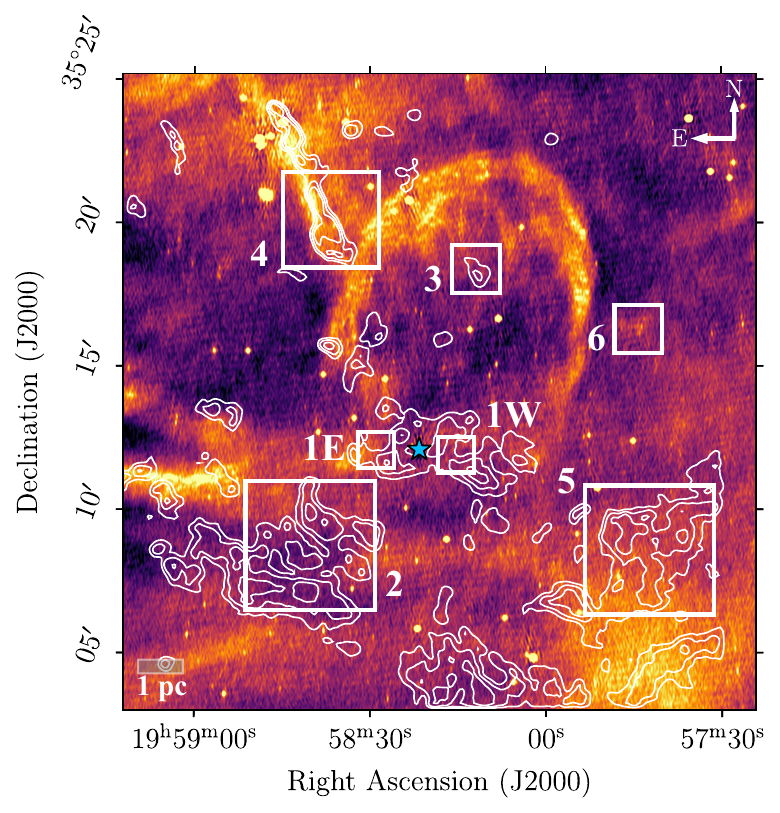}\hfill
    \includegraphics[width=0.35\textwidth]{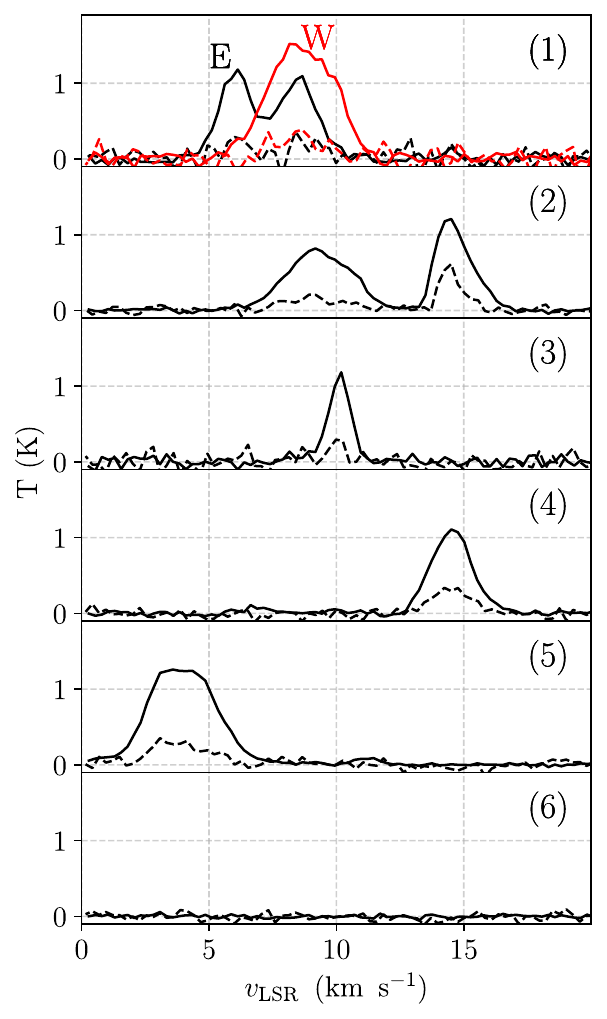}\hfill
    \caption{Spectral analysis of several regions of interest surrounding Cyg X--1. \textit{Left:} Radio continuum (MeerKAT) reference image with the IRAM--30m $^{12}$CO contours overlayed (contour levels [4.8, 9.6, 16.3] K km s$^{-1}$). The central BHXB source is marked by a blue star. Spectra from five regions of interest around this field have been extracted, in addition to an emission-free region to assess the noise level (white boxes). \textit{Right:} Spectra from the six highlighted regions in the \textit{left} panel. The solid lines represent the $^{12}$CO ($J=2-1$) intensity, and the dashed lines represent that of $^{13}$CO intensity (scaled by a factor of $\times3$ for visualization purposes). Regions 1 and 2 show multi-peaked spectral components. Regions 3 and 4 display mildly asymmetric line profiles. Region 5 shows a broad emission line profile, with clear velocity separation from the rest of the regions. Region 6 shows no line emission.} 
    \label{fig:spectra}
\end{figure*}

\begin{figure*}[t!]
    \centering
    \includegraphics[width=\textwidth]{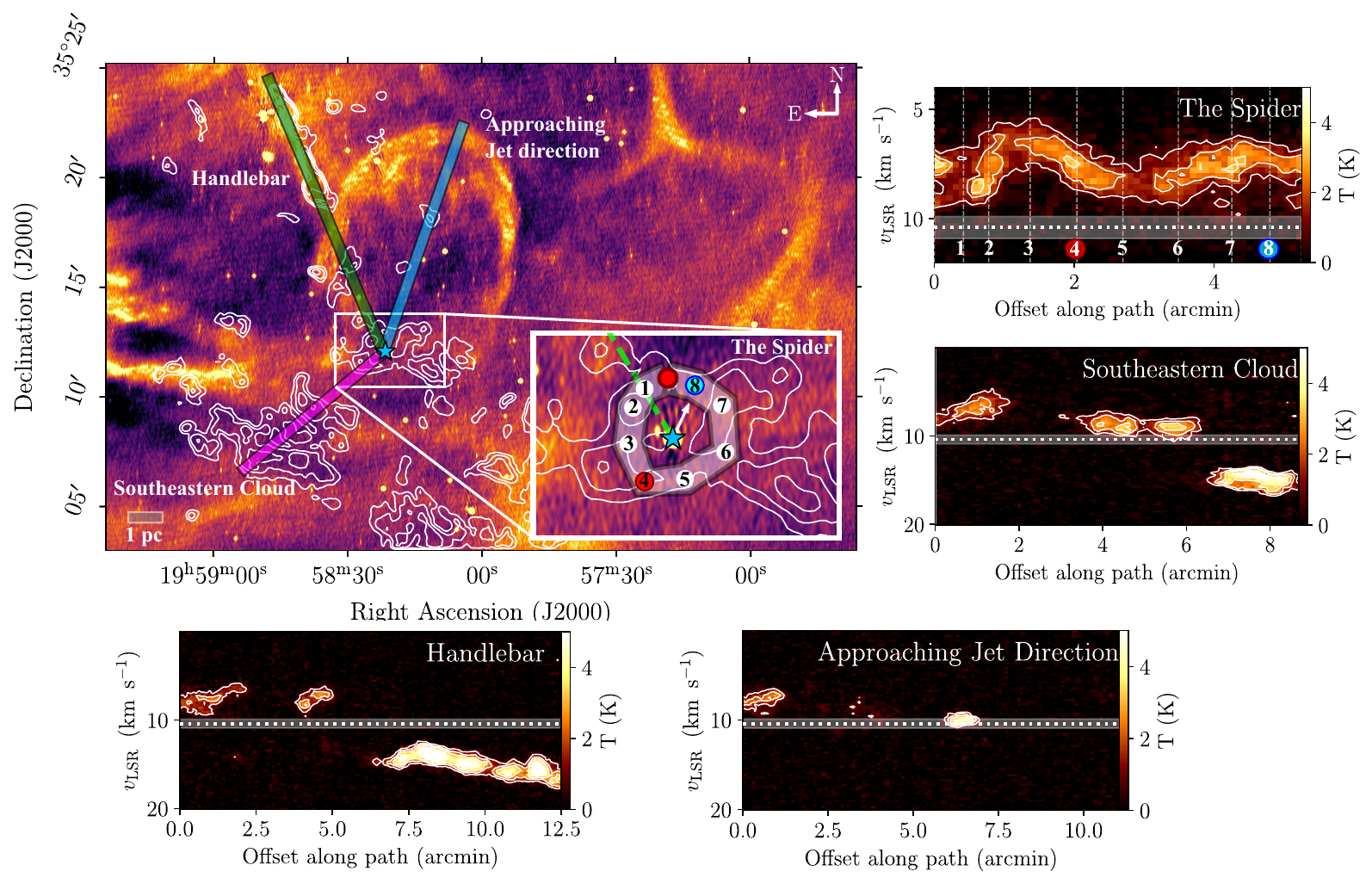}
    \caption{Kinematic analysis of $^{12}$CO($J=2-1$) emission in the Cyg X--1 field. \textit{Left:} Radio continuum (MeerKAT) reference image with the IRAM--30m $^{12}$CO contours overlayed (contour levels [4.8, 9.6, 16.3] K km s$^{-1}$). The BHXB position is marked with a blue star and the $30''$-wide P-V extraction paths are taken as follows: along paths starting from the central BHXB and directed towards the approaching jet (blue), \textit{the southeastern cloud} (magenta), and \textit{the handlebar} structure (green), as well as a ring around \textit{the spider} structure (white) starting from the red marker and proceeding counterclockwise. Within the inset, the white arrow represents the direction of the approaching jet and the green dashed line represents the past path of Cyg X--1's proper motion. \textit{Right \& below:} $^{12}$CO P-V diagrams along the slices indicated in the \textit{left} panel (contours at [1.0, 2.0, 3.0] K). The horizontal dotted line/gray shading represents the $v_{\text{LSR}}$ of Cyg X--1 with its uncertainty range. The \textit{top right} panel shows the position of the reference points around the \textit{the spider} complex with vertical dashed lines (these are also indicated as numbered dots in the inset image in the \textit{left} panel), and the red and blue circles around points 4 and 8 indicate the receding and approaching directions of the jet, respectively . \textit{The spider} complex shows an undulating velocity pattern, possibly pointing towards an expanding structure centered on the BHXB source. Displaced gas towards increasingly positive velocities (red-shifted) is evident in the \textit{southeastern cloud} direction, consistent with the action of a receding jet. No clear indication of an interaction with a jet is evident in the approaching jet direction or towards \textit{the handlebar} structure, as both P-V diagrams show red-shifted motion inconsistent with an approaching jet.}
    \label{fig:PV}
\end{figure*}

\section{Results} \label{sec:res}

\subsection{Molecular line emission morphology}
In Fig.~\ref{fig:momentmaps}, we show integrated intensity maps of $^{12}$CO (optically thick) and $^{13}$CO (optically thin, tracing denser gas) covering the Cyg X--1 region. These maps display significant (3$\sigma$) detections of extended emission regions for both molecular transitions. 

Upon first inspection, when compared to radio continuum and optical observations from previous analyses, we note that the main region inside and around the bow shock is mostly devoid of molecular gas within the probed velocity range (see the dash-dotted lines in Fig.~\ref{fig:comparison}; \citealt{atri2025quantifying,russell2007jet}). 

We also identify two new bright molecular emission regions near the BHXB (see Fig.~\ref{fig:momentmaps}, \textit{left}): (i) surrounding the BHXB in an irregular shape (called \textit{the spider} hereafter), and (ii) a bright, compact region just off the jet axis in the receding jet direction (called \textit{the southeastern cloud} hereafter) which shows an indentation-like morphology reminiscent of a bow shock (vertically-dashed green line on Fig.~\ref{fig:comparison}, {\it left}).
Lastly, we find several other molecular features located farther away from the central BHXB that will also be further inspected and analysed. Namely, there is an elongated feature north east of the source that is pointed towards the BHXB (called \textit{the handlebar} hereafter) and an extended cloud towards the south west direction (called \textit{southwestern cloud} hereafter).

\subsection{Molecular line spectral analysis}
After a first inspection of the integrated molecular emission, we proceeded to perform a spectral analysis by extracting  $^{12}$CO and $^{13}$CO region-averaged and unmasked spectra from the five previously highlighted regions, and a further sixth to assess the noise of an emission-free region (see Fig.~\ref{fig:spectra} \textit{left} for extraction regions, \textit{right} for spectra). 

We highlight the following features from the $^{12}$CO and $^{13}$CO spectra (Fig.~\ref{fig:spectra}, \textit{right}, solid lines and dashed lines, respectively):
\begin{itemize}
    \item \textbf{Regions 1E and 1W} cover the \textit{the spider}, where we compare different sections of this structure located to the east and west of the central BHXB. We find different spectral profiles in both of these regions. On the eastern side (1E), we identify a bi-modal $^{12}$CO line profile peaking at $\sim 6$ km s$^{-1}$ and $\sim 8$ km s$^{-1}$, with wide line profiles that span within the range $\sim [5,10]$ km s$^{-1}$. On the western side (1W), we find a much wider $^{12}$CO line profile $\sim[5,12]$ km s$^{-1}$ peaking around $\sim 7.5$ km s$^{-1}$. The $^{13}$CO features are much fainter, showing a very slight intensity increase (significant at $3\sigma$ as seen in Fig.~\ref{fig:momentmaps}, {\it right}) within the $^{12}$CO counterpart emission range. 

    \item \textbf{Region 2} covers {\it the southeastern cloud}. We identify two separate line features in this region. On the lower velocity end, we find a wide line profile peaking at $\sim 8$ km s$^{-1}$ and spanning $\sim[6,12]$ km s$^{-1}$. On the higher velocity end, we find asymmetric and slightly brighter line profile peaking at $\sim 14$ km s$^{-1}$. The latter line also has a pronounced $^{13}$CO counterpart. 
    
    \item \textbf{Region 3} covers a compact molecular cloud lying along the approaching jet axis in the direction of the known bow shock. We identify a narrow line profile peaking at $\sim 10$ km s$^{-1}$ in this region, with a faint $^{13}$CO counterpart.
    
    \item \textbf{Region 4} covers \textit{the handlebar} structure, located to the east of the bow shock. This region shows bright $^{12}$CO and $^{13}$CO lines peaking at $\sim 14$ km s$^{-1}$. Additionally, this region shows bright radio (see Fig.~\ref{fig:comparison}, \textit{left}) and infrared (see App.~\ref{sec:IR}) continuum emission. 
    
    \item \textbf{Region 5} covers the \textit{southwestern cloud} structure. In this region, we observe a very wide and flat-topped $^{12}$CO line profile at roughly $\sim 3-5$ km s$^{-1}$. The $^{13}$CO emission mimics the same behavior but at a lower intensity.
    
    \item \textbf{Region 6} covers an emission-free region to assess the noise contribution, showing no line-emission as expected.
\end{itemize} 

Overall, our spectral analysis reveals multi-peaked and asymmetric line profiles (particularly in Regions 1 and 2) in several regions around Cyg X--1. These emission line properties are similar to past work results \citep{tetarenko2018mapping, tetarenko2020jet, bosch2026constraining}, showing evidence of dynamical interactions in previous applications of this astrochemical technique around BHXBs. Additionally, we find that the densest environments are Regions 2, 4, and 5, where $^{13}$CO has stronger emission intensity levels.

\subsection{Molecular line kinematic analysis}

Following from our spectral analysis, we examined the gas kinematics in several regions around Cyg X--1. Namely, we analyze $30''$-wide paths near regions 1 (\textit{the spider}), 2 (\textit{southeastern cloud}), 3 (approaching jet direction), and 4 (\textit{the handlebar}). Fig.~\ref{fig:PV} shows $^{12}$CO Position-Velocity (P-V) diagrams exploring gas kinematics along the aforementioned paths. We do not perform this kinematic analysis on Region 5, as its lack of asymmetric or multi-peaked line features and its velocity peak being so different to the rest of the regions and that of Cyg X--1 suggest it is unrelated to the feedback from the central BHXB source.

Around {\textit{the spider}}, we find an undulating velocity structure with a median velocity of $\sim7.5$ km s$^{-1}$ (see Fig.~\ref{fig:PV}). The maximum velocity difference between peaks is $\leq 5$ km s$^{-1}$. All features seem to be consistently blueshifted with respect to the velocity of Cyg X--1\footnote{This blueshift could be due to a combination of opacity and geometry effects that cannot be disentangled with this observation alone.}. The spider shows strong velocity gradients, of up to $\sim 3~\text{km s}^{-1}~\text{arcmin}^{-1}$, which suggest strong local perturbations with rapid spatial variations. These strong gradients are unique to this region (see App.\ref{sec:circ}), suggesting a possible outflow-ISM interaction at this site.

In \textit{the southeastern cloud}'s direction we find progressively red-shifted clouds of gas as we move farther away from the central BHXB source, with a total velocity displacement of $\sim7$ km s$^{-1}$ between observed components. The southeastern cloud is split into two components with different bulk motions with respect to Cyg X--1: (i) a closer, clumpy slightly blueshifted component, and (ii) a further, brighter component showing marginally increasing redshifted velocities as the distance to the source grows. 

In \textit{the approaching jet direction}, we also observe marginally increasing red-shifted velocities in isolated molecular clouds as we move away from the central BHXB source. In particular, the furthermost region shows a velocity fully consistent with that of Cyg X--1. This kinematic structure shows no indication of the action of an approaching jet, which would display some blue-shifting behavior.

In {\it the handlebar}'s direction, we find a bright and progressively red-shifted structure as we move away from the central BHXB source, with a velocity gradient of $\sim 0.8~\text{km s}^{-1}~\text{arcmin}^{-1}$. These kinematics are inconsistent with an approaching jet-blown cocoon colliding with such a gas structure. Furthermore, the detected continuum infrared emission (discussed in App.~\ref{sec:IR}) and the apparent connection with the Tulip nebula H\textsc{ii} region through a continuous filament visible in the CO emission (Fig.~\ref{fig:momentmaps}), radio emission (Fig.~\ref{fig:comparison}) and IR emission (App.~\ref{sec:IR}) could suggest that this region is unlikely to be driven by a Cyg X--1 based outflow.

Overall, our P-V diagrams provide evidence that points towards possible interactions between Cyg X--1 based outflows and the ISM in the \textit{the spider} and in \textit{the southeastern cloud}. However, in the approaching jet direction and in \textit{the handlebar} structure, we do not find such indicators. 
In the discussion below, we will focus on \textit{the spider} complex and \textit{the southeastern cloud}, analyzing the implications of these observations and attempting to reconcile them with outflow interaction models.

\section{Discussion} \label{sec:disc}


We identified an extended, irregular molecular structure (\textit{the spider}) around Cyg X--1 with complex kinematics (including the absence of a clear blue/redshift pattern aligned with the jet axis or the proper motion path of Cyg X--1) suggestive of an interaction between multiple different outflows and the ambient ISM. 
The nature of this interaction depends critically on the relative three-dimensional velocity between Cyg X--1 and the ambient ISM\footnote{We note that our gas kinematics constraints are based on the P-V diagrams alone. However, the path extracted in our P-V diagrams may not fully capture the intrinsic three-dimensional geometry of the gas structures, and the optical thickness of our tracer $^{12}$CO may bias our view toward the near side of the structure.}. In the following sections, we explore two different scenarios to explain the origin of \textit{the spider}.


\subsection{High relative motion between Cygnus X--1 and the ambient ISM} \label{sec:runaway}

\begin{figure*}
    \centering
    \includegraphics[width=0.48\textwidth]{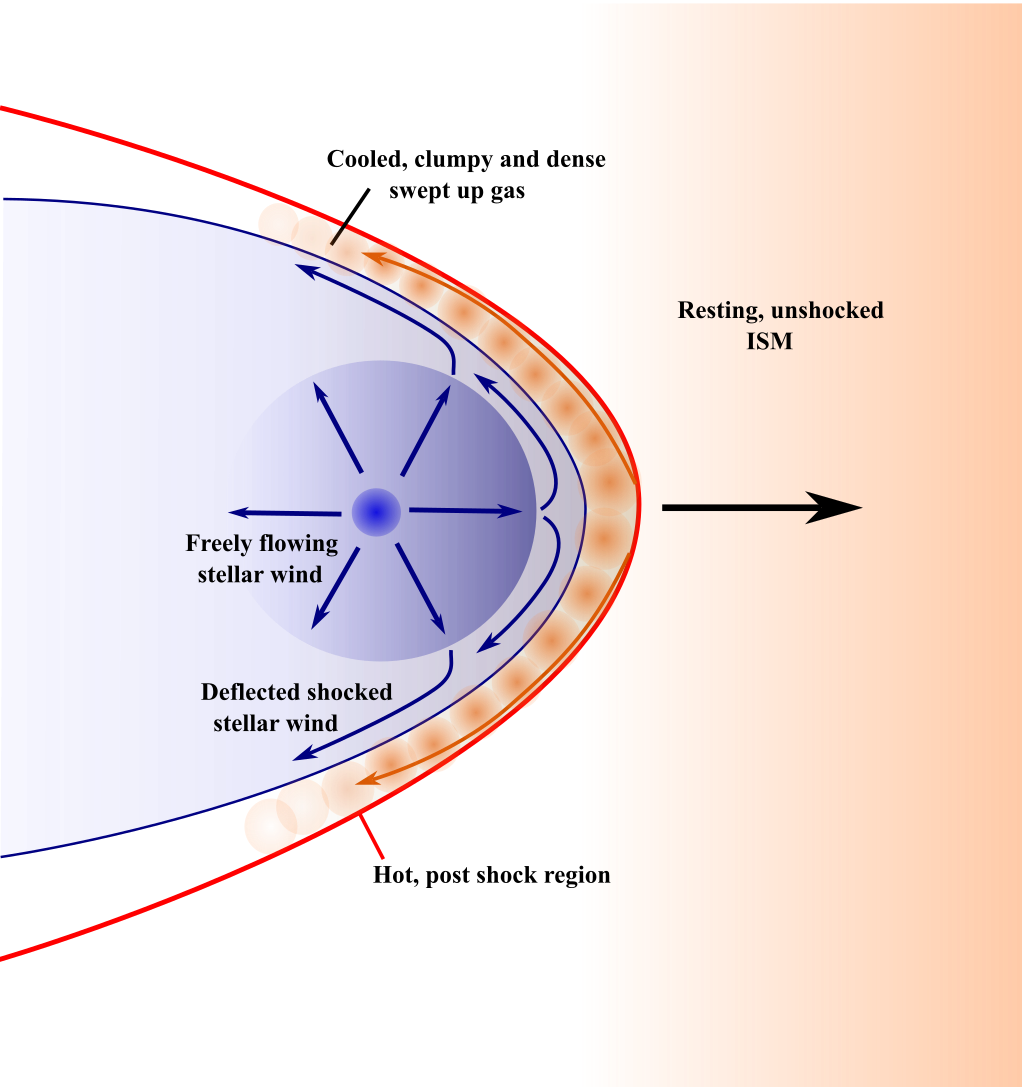}
    \includegraphics[width=0.48\textwidth]{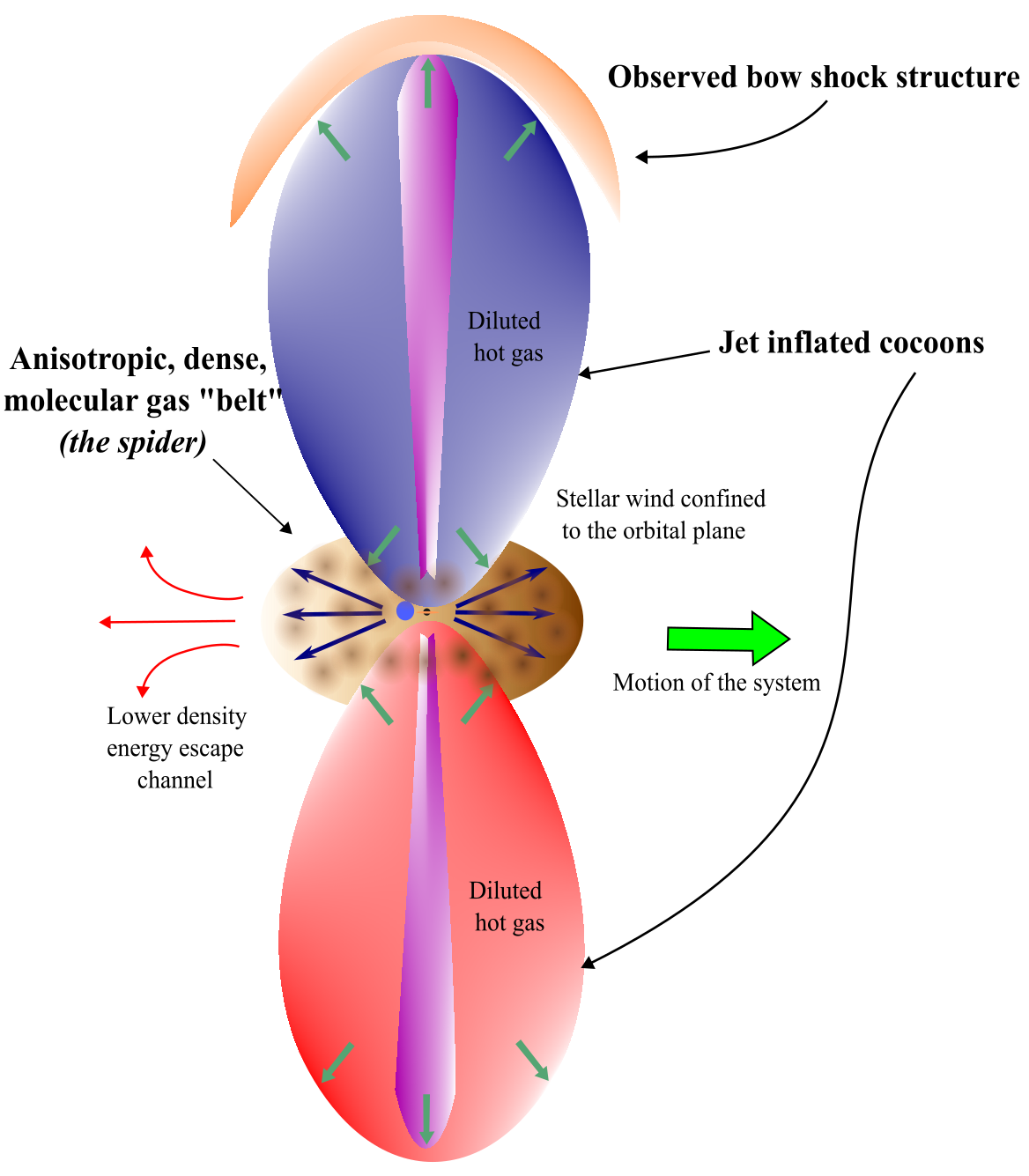}
    \caption{Schematics representing the main elements of the scenario proposed in \S\ref{sec:runaway}. We represent Cyg X--1 at the center and moving towards the direction of the green arrow (right side of the page). Initially the star creates a bow shock that sweeps up material, cooling it effectively and creating a layer of clumpy and dense molecular gas between the hot leading edge of the shocked ISM and the hot shocked wind (brown bow, \textit{left panel}). As the wind strengthens and the jets turn on in a new powerful phase, they inflate cocoons (blue for the approaching jet component and red for the receding jet component) that confine the stellar wind to the orbital plane, reshaping the molecular gas into an anisotropic belt with complex velocity structure (brown ellipsoid, \textit{the spider, right}). We note that wind-ISM interactions in bow shocks are momentum-driven because a lower density channel forms in the wake of the system's path as it moves through the ISM, allowing for the shocked wind to escape (blue arrows on the {\it left} panel, red arrows on the {\it right} panel).  We note that the approaching and receding jet blown cocoons may be asymmetric but we opted for a simple representation in this diagram. 
    None of the elements in these schematics are set to scale, and the offset angle between the motion and jet axis is taken to be perpendicular for a simplistic assessment.} 
    \label{fig:model}
\end{figure*}

Massive stars moving with high velocities relative to the ambient ISM are known to form bow shocks in their direction of motion \citep[see Fig.~\ref{fig:model}, {\it left}; e.g.,][]{comeron1998numerical, martinez2023probing}. In the situation where Cyg X--1 has a relative velocity of $10-20~\text{km s}^{-1}$ with respect to the local ISM, the formation of such a bow shock would be expected. Within these structures, as the system moves with respect to the ambient ISM, the swept material can be radiatively cooled to form dense and clumpy molecular gas at the interface between the shocked stellar wind and the leading edge of the bow shock. The cooling rate is efficient and rapid given the post-shock temperatures of $T_s\sim10^4~\text{K}$ \citep[for a shock moving at $10-20$ km s$^{-1}$;][]{dopita2003astrophysics}, and the shock compression of the ISM can yield densities an order of magnitude higher than those present in the ambient, unperturbed ISM \citep[as seen in numerical simulations, e.g.,][]{meyer2014models, acreman2016modelling}. In turn, the cooling rate can be estimated as:
\begin{equation}
    t_{\rm cool}=\frac{E}{\dot{E}}=\frac{\frac{3}{2}n_sk_BT_s}{n_s^2\Lambda(T_s)}
\end{equation}
Where we can take a shock number density of $n_s\sim10-100~\text{cm}^{-3}$ for an ambient gas of initial density $n_{\rm ISM}\sim1-10~\text{cm}^{-3}$, and a cooling rate of $\Lambda(T)\sim10^{-22}-10^{-23}~\text{erg cm}^3~\text{s}^{-1}$ \citep{sutherland1993cooling}\footnote{Assuming a gas of solar metallicity, in absence of a better estimate.}. Depending on the exact parameters of the initial configuration, the cooling timescale will take values of $t_{\rm cool}\sim10^1-10^3~{\rm years}$, well below the relevant dynamical evolution timescales of this structure. Therefore, the formation of a structure surrounding the star, cool enough to host significant amounts of molecular gas and in approximate pressure equilibrium with the shocked wind, is naturally explained by this scenario.

Accounting for the small radial velocity difference between Cyg X--1 and the median velocity of the observed region, the bow shock should display velocities close to that of the system in the direction of motion and velocities more similar to the ambient ISM elsewhere \citep{martinez2022nonthermal}. However, the clumpy nature of molecular gas combined with opacity effects could cause deviations from this behavior, giving rise to more intricate velocity structures. 

Assuming a velocity of the system within the ambient ISM ($v\sim10-20~\text{km s}^{-1}$) and the observed size of {\it the spider} ($R_c\sim2~\text{pc}$), it would take at least $t_f\sim R_c/v\sim10^5~\text{yr}$ to form {\it the spider} to its current size, comparable to the lifetime of the system\footnote{While there is a difference between the estimated age of \textit{the spider} and that of the jet blown cocoons ($\leq 4\times 10^4~\text{yr}$, \citealt{russell2007jet,sell2015shell}), we estimated the former in a very simplistic order of magnitude assessment, so the two may in fact be similar.}.

With the existence of the known cocoon structure to the north of Cyg X--1, which is presumably inflated by the approaching jet \citep{gallo2005dark,russell2007jet,sell2015shell}, we also need to consider whether jet activity may also have played a role in shaping \textit{the spider}. The current phase of the jet activity in Cyg X--1 may have been triggered by changes in the accretion dynamics of the system, which are likely to also affect the stellar wind. It is therefore likely that \textit{the spider} structure could be a result of a coupled jet-stellar wind interaction evolving while the source moves within a relatively dense ambient medium. However, the detailed observational imprint of such complex dynamics is hard to predict without carrying out hydrodynamical simulations, which are beyond the scope of this work. 

Prior to the significant jet activity in Cyg X--1, the stellar wind likely produced a modest bow shock ($\lesssim$pc scale; \citealt{martinez2022nonthermal}) in the direction of the system's motion with respect to the ISM. Since Cyg X--1 is thought to have formed with a very weak or non-existent supernova explosion \citep{mirabel2003formation}, and the system has a known proper motion with respect to its surrounding stars \citep{carretero2025observational, nagarajan2025mixed}, the current surrounding medium should have been clear of any significant remnant structures associated to a previous evolutionary stage of Cyg X--1.

As the stellar companion evolves, the stellar mass loss rate and the Roche lobe filling factor increase, triggering changes in the accretion dynamics of the system by increasing the intensity of the stellar wind, and presumably the power of the jets \citep{hirai2021conditions}. When the stellar wind strengthens, it drives a stronger shock into the ambient medium, sweeping up material over scales of a few parsecs (a few arcmin at the distance of Cyg X-1, as we observe). At the same time, the jets inflate bipolar cocoons that partially confine the wind-driven bow shock to a belt around the orbital plane (see Fig.~\ref{fig:model}, {\it right}). This interaction leads to an anisotropic structure: while the wind pushes and accumulates shocked ISM mainly upstream of the system motion, the shocked wind can escape downstream through a low-density channel associated with the original bow shock \citep{bosch2011termination}. This energy escape channel suggests that the interaction is momentum-driven since the stellar wind energy mostly escapes downstream with the hot wind material \citep{martinez2022nonthermal}, and it can not accumulate to dominate the evolution of the structure.
The structure we identify as \textit{the spider} could therefore be a dense, cooled, belt-shaped, anisotropic ISM shell produced by the motion-driven bow shock and the presence of the jet cocoons, confining the wind to the jet axis perpendicular plane. The dense nature of the radiative material accumulated at the edge of \textit{the spider} is further supported by the critical density associated to the $^{12}$CO$(J=2-1)$ transition ($n_{crit}\sim10^3$ cm$^{-3}$; \citealt{komugi2025alma}), explaining the emission we observe.

We note that we are describing a simplified version of this scenario, since we are taking the jet axis to be perpendicular to the direction of motion. The plane where the spider forms and the direction of the energy escape channel presented in Fig .~\ref{fig:model} may not be of $90^\circ$. Existing measurements suggest an angle between the systems motion direction and jet axis of $\sim110-120^\circ$ \citep{miller2021cygnus,nagarajan2025mixed}, however, local dynamics of the ambient ISM may differ from this estimate. A possible additional offset introduces deviations from a simplistic picture that require detailed relativistic hydrodynamical simulations to assess, which fall out of the scope of this work. Instead, we will keep this first-order approach to evaluate some semi-quantitative indicators of the viability of this physical situation.

A consistency test of this scenario is to compare the observed radiative output of the cold molecular gas to the expected energy flux (supplied by the wind through ram pressure) remaining after the shocked ISM has cooled down to molecular gas temperatures.
The molecular gas radiative output should, in principle, account for a fraction of the kinetic luminosity supplied by the wind through ram pressure, given the efficient radiative cooling mechanisms that molecular gas typically exhibits, but not exceed this value, given the possible presence of other coolants.

The energy flux supplied by the wind and available to be radiated away in the cooled gas of \textit{the spider}, can be approximated as $L_{\rm spider} \sim S~c_s~P_{\rm ram}$, where $S$ is the interaction surface, $c_s$ the sound speed of the molecular gas, and $P_{\rm ram}\sim\rho_w v_w^2$ the wind ram pressure at the characteristic radius. We adopt simple geometric and physical assumptions to estimate this quantity to order of magnitude.
We approximate the interaction surface as a fraction of a sphere, $S \sim 4\pi r^2 f$, with $r\sim2$ pc and $f\lesssim 1$, consistent with the size and morphology of \textit{the spider}. For the cold molecular gas, we adopt a sound speed $c_s \sim 0.4$–$0.8$ km s$^{-1}$, corresponding to temperatures of $50$–$150$ K \citep[as typically found in molecular gas shocked by protostellar object outflows; e.g.,][]{van2009dense}. Using the sound speed of the cold phase provides an estimate of the energy transfer rate.
The wind ram pressure is estimated assuming standard O-star wind profiles \citep{puls2008mass},
\begin{equation}
     \rho(r)=\frac{\dot{M}}{4\pi r^2 ~v_\infty};\quad \text{for~}r>>R_* ,
\end{equation}
where $v_\infty$ is the terminal velocity of the stellar wind \citep[$\sim1600-2100~\text{km s}^{-1}$;][]{herrero1995fundamental, gies2008stellar}, $\dot{M}$ is the mass loss rate \citep[$\sim10^{-6}~\text{M}_\odot~\text{yr}^{-1};$][]{gies2003wind}, and $R_*$ is the radius of the O-star. For the luminosity, we obtain
\begin{equation}
    L_{\text{spider}}\gtrsim S~c_s~\rho(r)~v_\infty^2\sim\dot{M}~v_\infty~f~c_s.
    \label{eq:ramp}
\end{equation}
This yields $L_{\text{spider}} \sim 10^{32}$ erg s$^{-1}$. A fraction of this energy is expected to be radiated away by the molecular gas. From the observed CO($J=2-1$) emission, we estimate a luminosity $L_{\rm CO}\sim3\times10^{31}$ erg s$^{-1}$ (using the Rayleigh-Jeans approximation), which is lower than $L_{\rm spider}$, and consistent with a radiatively efficient, momentum-driven interaction.

In contrast, regions impacted by the jets directly experience much higher shock velocities, leading to hotter radiative gas (that can be traced by optical line emission and radio continuum). This naturally produces a multiphase structure, where the cold molecular component associated mostly to the stellar wind-ISM interaction traces \textit{the spider}, while hotter gas is associated with the jet-inflated cocoons. However, as discussed in \S\ref{sec:obs}, the presence of molecular gas near the edge of the cocoon can not be fully ruled out, since the velocity range observed does not cover the expansion velocity of the bow shock's leading edge.

Overall, this scenario (see Fig.~\ref{fig:model}) explains the coexistence of a slower, dense molecular structure near the system, driven primarily by the motion of the system and the stellar wind interacting with the ISM ({\it the spider}), and a more extended, higher-temperature, higher-velocity, jet-driven cocoon (giving rise to the observed bow shock towards the NW of the source, \citealt{gallo2005dark, russell2007jet}).

\subsection{Low relative motion between Cygnus X--1 and the ambient ISM} \label{sec:rest}

Given that Cyg X--1 lies close to star forming regions, the possibility that the ambient ISM could be perturbed by the injection of turbulent motions and large-scale bubbles exists. In this scenario, the perturbation could be such that the ambient ISM velocity would match that of Cyg X--1, down to fluctuations of a few km s$^{-1}$. In this limit, the interaction between the stellar wind and the ISM is expected to be energy-driven\footnote{We discard an SNR-like momentum-driven radiative ``snow-plow" regime, as in this case, the wind injects energy continuously. Even if cooling causes a shell to become thinner and denser, looking similar to a ``snow-plow" regime to order zero, the evolution of the structure would energy-driven.}, as the shocked wind would remain confined and inflate a hot bubble \citep{castor1975interstellar}.

To assess the expected dynamical impact of this scenario, we consider the amount of energy injected by the wind over the lifetime of the jet-inflated cocoon, $\lesssim 4\times10^4$ yr \citep{russell2007jet, sell2015shell}, which provides a direct constraint on the recent interaction timescale in the system. The kinetic luminosity of the stellar wind,
\begin{equation}
L_w = \frac{1}{2}\dot{M}v_\infty^2 \sim 10^{36}~\mathrm{erg~s^{-1}},
\end{equation}
then implies a total injected energy of $E_w \sim 10^{47}-10^{48}$ erg over $\sim 4\times10^4$ yr. If efficiently converted into bulk motion, this would yield expansion velocities of order $\sim 100~\text{km s}^{-1}$, well above the observed values.

Reconciling the high velocities expected from an energy-driven bubble with the low observed velocities of {\it the spider} requires a different configuration. If instead {\it the spider} is a foreground molecular structure, it could be impacted by the expanding wind-driven bubble and accelerated through a momentum-driven interaction.

In this case, the acceleration timescale for the stellar wind bubble to accelerate the molecular cloud can be estimated as
\begin{equation}
t_a \sim \frac{v_c}{a}\sim\frac{M_c v_c}{\epsilon \dot{M} v_\infty},
\end{equation}
where $v_c$ is the cloud velocity, $a$ is the cloud acceleration\footnote{Estimated through the momentum rate (force) of the wind (i.e., $F=\dot{p}\sim\dot{M}~v_\infty; \quad a=F/M_c$).}, $M_c$ is the cloud mass, and $\epsilon$ parametrizes the efficiency of momentum transfer between the wind and the molecular gas. Taking $M_c \sim 10~\mathrm{M}_\odot$ (through H$_2$ column density map, see App.\ref{app:columndens}), $v_c \sim 5~\mathrm{km~s^{-1}}$, and $\epsilon \sim 0.1$--$1$ yields $t_a \sim 10^4$--$10^5$ yr.

This timescale is short compared to the $\sim$Myr lifetime of the system \citep{miller2021cygnus}, raising the question of why no clear signatures of cumulative wind-ISM interaction are observed from earlier epochs. In addition, this scenario requires both a small relative velocity between Cyg X--1 and the ambient ISM and a favorable line-of-sight configuration in which a dense cloud is located in the foreground and efficiently coupled to the wind-driven flow.

Given the amount of finely-tuned conditions necessary for this scenario to be viable, we conclude that the interaction discussed in \S\ref{sec:runaway} provides a more natural explanation for the origin of {\it the spider}, if related at all to Cyg X--1.

\subsection{Probability of association between Cygnus X--1 and the spider} \label{sec:nonassoc}
Given the prevalence of molecular emission within the observed region, and the lack of strong evidence for a significant part of it to be related to Cyg X--1, it is useful to assess the probability that \textit{the spider} is indeed related to Cyg X--1. 

We can start by assessing what fraction of the image is occupied by molecular emission. We compute this by taking all of the 3$\sigma$ $^{12}$CO emission (see Fig.~\ref{fig:momentmaps}, \textit{left}), which covers roughly 33\% of the image.
However, this is a broad estimate, given that we are not taking into account any velocity restrictions. If instead, we select only the emission within the Cyg X--1 velocity range\footnote{We take the range $v_{\rm LSR}=(0.7,10.7)~\text{km s}^{-1}$. Note that this selected emission includes the  emission from \textit{the spider}.} (including uncertainties), then the probability of \textit{the spider} being a random molecular cloud that happens to be spatially coincident with Cyg X--1 is reduced to a 5\%.

As another consistency check for the uniqueness of \textit{the spider}, we have explored circular paths around the different masses of gas present in the field (see App.\ref{sec:circ}, Fig.~\ref{fig:PVcirc}) to look for similar velocity structures. While some undulation naturally appears on a circular path covering a region with slightly different velocity peaks, we find that the amplitudes of the velocity peaks and the velocity gradients of \textit{the spider} are clearly distinct from these other regions.

Therefore, we conclude that given the relatively low chance alignment probabilities and the uniqueness of the velocity structure in \textit{the spider}, an association between Cyg X--1 and \textit{the spider} is likely.

\subsection{The origin of the southeastern molecular cloud} \label{sec:reced}

We identify another extended molecular structure south east from the central BHXB source and consider whether it may be consistent with gas being pushed away from Cyg X--1, possibly by the receding jet. 
This component (spatially corresponding to Region C in Fig.~\ref{fig:columndensity} and Fig.~\ref{fig:PV}, \textit{Southeastern Cloud panel}) shows a consistent redshift with respect to Cyg X--1, with a slight velocity gradient ($\sim0.7~\text{km s}^{-1}~\text{arcmin}^{-1}$), showing increasingly redshifted velocities as the distance from the source grows. While the kinematics pattern of this region might suggest a receding motion, there is a spatial offset of $\sim25^\circ$ with the known jet axis. Stellar-wind driven jet bending is known to widen the initial jet opening angle ($\sim0.8^\circ$; \citealt{miller2021cygnus}), but this interaction only affects the system at very close scales, and its large scale effects should be negligible in the case of Cyg X--1 \citep{prabu2026jet}. The observed $\sim25^\circ$ indicates that this region cannot be directly affected by the jet's flow. Instead, if the receding jet has also inflated a cocoon (similar to what is observed for the approaching jet to the north), this region may exist within its boundaries, and effectively showing receding motion as well. We note, however, that the receding velocity of the cloud with respect to Cyg X--1 is significantly slower than the expected expansion velocity of the cocoon \citep[$>90~\text{km s}^{-1}$][]{russell2007jet, sell2015shell}. However, we note that the dynamics of such a region when interacting with strong shocks could be very complex and would require hydrodynamical simulations to understand the possible velocities expected within dense phases of the ISM.

An indication of the existence of a receding jet-driven cocoon has long been sought after \citep{russell2007jet, atri2025quantifying}, given the evident cocoon and related bow shock found north of the source. The location of Cyg X--1 with respect to the Tulip Nebula (north east from the source) partially explains why the mass sweeping of a cocoon could form a visible bow-shock in the approaching direction and fail to do so in the receding direction. However, the existence of this southeastern cloud provides an intriguing prospect for a potential receding jet driven interaction that should be investigated with future higher spatial resolution molecular line observations of the region, and further shock and temperature tracers, which can yield additional evidence for a jet-driven interaction \citep[e.g. with NOEMA,][]{tetarenko2018mapping}.

\section{Summary}

In this work, we investigate the molecular ISM in the local environment of the BHXB Cyg X--1 to search for outflow-driven interaction signatures. Our IRAM--30m telescope observations reveal bright and extended molecular gas emission in the surveyed region.
However, there is no significant molecular emission in the direction of the known radio continuum/optical line-bright bow shock within the $v_{LSR}=[-100,100]~\text{km s}^{-1}$ range.

We find two molecular features potentially driven by outflow-based interactions, which have not been identified in previous studies.
In particular, we find a structure surrounding the BHXB (\textit{the spider}) with an undulating velocity pattern revealed by the P-V diagrams (see Fig.~\ref{fig:PV}, \textit{top right}). We find that this structure may be consistent with an interaction driven by the stellar wind from the companion star. The favored scenario we propose suggests that the stellar wind, together with the star's peculiar velocity with respect to the local ISM, could have created a small bow shock that accumulated radiatively cooled, dense material at its edges. This bow shock may have grown due to an increase in the intensity of the stellar wind, and further modified by the current powerful phase of the relativistic jets and laterally compressed by the cocoons they inflate through redirecting the stellar wind towards the orbital plane (shown in the schematic of Fig.~\ref{fig:model}). We have also explored further scenarios based on the relative position and velocities between Cyg X--1, {\it the spider}, and the ambient ISM, but find the above scenario to be the most natural explanation given the nature of Cyg X--1. A well traced, higher spatial resolution \textit{spider} velocity structure could provide in this scenario important information on the medium-binary-jet relation. This would be an interesting observational goal to pursue in the future.

We also identify a secondary structure south east from the source, which shows recessional motion away from the central BHXB. While these kinematics could be explained by the receding jet driving dense clouds of gas away from Cyg X--1, the large spatial offset between this cloud and the known jet axis may suggest that the two are in fact not related. However, we note that further observational and theoretical work is needed to better understand the possible connection between a jet cocoon and this dense instance of gas within this complex region.

Overall, our results validate the potential of molecular line imaging to identify and characterize outflow-ISM interaction sites near BHXBs and potentially help distinguish between multiple sources of feedback (in the case of Cyg X--1, a relativistic jet and an O-star stellar wind).

\begin{acknowledgments}
The authors thank the anonymous reviewer for their insightful comments, which have improved the quality of this manuscript. PB--C and AJT acknowledge that this research was undertaken thanks to funding from the Canada Research Chairs Program and the support of the Natural Sciences and Engineering Research Council of Canada (NSERC; funding reference number RGPIN--2024--04458). VB-R acknowledges financial support from the State Agency for Research of the Spanish Ministry of Science and Innovation under grants PID2022-136828NB-C41 and CEX2024-001451-M funded by MICIU/AEI/10.13039/501100011033/ERDF/EU. VB-R is a Correspondent Researcher of CONICET, Argentina, at the IAR. DMR is supported by Tamkeen under the NYU Abu Dhabi Research Institute grant CASS. SP acknowledges Breakthrough Listen (funded by the Breakthrough Prize foundation) for their Breakthrough Listen fellowship. This work is based on observations carried out under project number 058--18 with the IRAM--30m telescope. IRAM is supported by INSU/CNRS (France), MPG (Germany), and IGN (Spain). We additionally thank the IRAM staff for their efforts carrying out the observations presented in this paper. The reduction of the presented datasets was performed using the \textsc{Gildas} software \citep{gildas}. This work made use of the \textsc{carta} (Cube Analysis and Rendering Tool for Astronomy) software \citep[][–https://cartavis.github.io]{CARTA}. This research has made use of the NASA/IPAC Infrared Science Archive \citep{irsa1}, which is funded by the National Aeronautics and Space Administration and operated by the California Institute of Technology.
\end{acknowledgments}

\facilities{IRAM--30m telescope, MeerKAT, ING:Newton, IRSA, WISE.}

\software{\textsc{Gildas} \citep{gildas}, astropy \citep{2013A&A...558A..33A,2018AJ....156..123A,2022ApJ...935..167A},
    PHANGS--ALMA pipeline \citep{leroy2021phangs},
    \textsc{SpectralCube} \citep{2019zndo...2573901G},
    \textsc{PVextractor} \citep{2016ascl.soft08010G}, \textsc{radex} \citep{van2007computer}, \textsc{Carta} \citep{CARTA}. }


\bibliography{sample701}{}

@ARTICLE{2022ApJ...935..167A,
       author = {{Astropy Collaboration} and {Price-Whelan}, Adrian M. and {Lim}, Pey Lian and {Earl}, Nicholas and {Starkman}, Nathaniel and {Bradley}, Larry and {Shupe}, David L. and {Patil}, Aarya A. and {Corrales}, Lia and {Brasseur}, C.~E. and {N{\"o}the}, Maximilian and {Donath}, Axel and {Tollerud}, Erik and {Morris}, Brett M. and {Ginsburg}, Adam and {Vaher}, Eero and {Weaver}, Benjamin A. and {Tocknell}, James and {Jamieson}, William and {van Kerkwijk}, Marten H. and {Robitaille}, Thomas P. and {Merry}, Bruce and {Bachetti}, Matteo and {G{\"u}nther}, H. Moritz and {Aldcroft}, Thomas L. and {Alvarado-Montes}, Jaime A. and {Archibald}, Anne M. and {B{\'o}di}, Attila and {Bapat}, Shreyas and {Barentsen}, Geert and {Baz{\'a}n}, Juanjo and {Biswas}, Manish and {Boquien}, M{\'e}d{\'e}ric and {Burke}, D.~J. and {Cara}, Daria and {Cara}, Mihai and {Conroy}, Kyle E. and {Conseil}, Simon and {Craig}, Matthew W. and {Cross}, Robert M. and {Cruz}, Kelle L. and {D'Eugenio}, Francesco and {Dencheva}, Nadia and {Devillepoix}, Hadrien A.~R. and {Dietrich}, J{\"o}rg P. and {Eigenbrot}, Arthur Davis and {Erben}, Thomas and {Ferreira}, Leonardo and {Foreman-Mackey}, Daniel and {Fox}, Ryan and {Freij}, Nabil and {Garg}, Suyog and {Geda}, Robel and {Glattly}, Lauren and {Gondhalekar}, Yash and {Gordon}, Karl D. and {Grant}, David and {Greenfield}, Perry and {Groener}, Austen M. and {Guest}, Steve and {Gurovich}, Sebastian and {Handberg}, Rasmus and {Hart}, Akeem and {Hatfield-Dodds}, Zac and {Homeier}, Derek and {Hosseinzadeh}, Griffin and {Jenness}, Tim and {Jones}, Craig K. and {Joseph}, Prajwel and {Kalmbach}, J. Bryce and {Karamehmetoglu}, Emir and {Ka{\l}uszy{\'n}ski}, Miko{\l}aj and {Kelley}, Michael S.~P. and {Kern}, Nicholas and {Kerzendorf}, Wolfgang E. and {Koch}, Eric W. and {Kulumani}, Shankar and {Lee}, Antony and {Ly}, Chun and {Ma}, Zhiyuan and {MacBride}, Conor and {Maljaars}, Jakob M. and {Muna}, Demitri and {Murphy}, N.~A. and {Norman}, Henrik and {O'Steen}, Richard and {Oman}, Kyle A. and {Pacifici}, Camilla and {Pascual}, Sergio and {Pascual-Granado}, J. and {Patil}, Rohit R. and {Perren}, Gabriel I. and {Pickering}, Timothy E. and {Rastogi}, Tanuj and {Roulston}, Benjamin R. and {Ryan}, Daniel F. and {Rykoff}, Eli S. and {Sabater}, Jose and {Sakurikar}, Parikshit and {Salgado}, Jes{\'u}s and {Sanghi}, Aniket and {Saunders}, Nicholas and {Savchenko}, Volodymyr and {Schwardt}, Ludwig and {Seifert-Eckert}, Michael and {Shih}, Albert Y. and {Jain}, Anany Shrey and {Shukla}, Gyanendra and {Sick}, Jonathan and {Simpson}, Chris and {Singanamalla}, Sudheesh and {Singer}, Leo P. and {Singhal}, Jaladh and {Sinha}, Manodeep and {Sip{\H{o}}cz}, Brigitta M. and {Spitler}, Lee R. and {Stansby}, David and {Streicher}, Ole and {{\v{S}}umak}, Jani and {Swinbank}, John D. and {Taranu}, Dan S. and {Tewary}, Nikita and {Tremblay}, Grant R. and {de Val-Borro}, Miguel and {Van Kooten}, Samuel J. and {Vasovi{\'c}}, Zlatan and {Verma}, Shresth and {de Miranda Cardoso}, Jos{\'e} Vin{\'\i}cius and {Williams}, Peter K.~G. and {Wilson}, Tom J. and {Winkel}, Benjamin and {Wood-Vasey}, W.~M. and {Xue}, Rui and {Yoachim}, Peter and {Zhang}, Chen and {Zonca}, Andrea and {Astropy Project Contributors}},
        title = "{The Astropy Project: Sustaining and Growing a Community-oriented Open-source Project and the Latest Major Release (v5.0) of the Core Package}",
      journal = {\apj},
         year = 2022,
        month = aug,
       volume = {935},
       number = {2},
          eid = {167},
        pages = {167},
          doi = {10.3847/1538-4357/ac7c74},
archivePrefix = {arXiv},
       eprint = {2206.14220},
 primaryClass = {astro-ph.IM},
       adsurl = {https://ui.adsabs.harvard.edu/abs/2022ApJ...935..167A}
}

@ARTICLE{2018AJ....156..123A,
       author = {{Astropy Collaboration} and {Price-Whelan}, A.~M. and {Sip{\H{o}}cz}, B.~M. and {G{\"u}nther}, H.~M. and {Lim}, P.~L. and {Crawford}, S.~M. and {Conseil}, S. and {Shupe}, D.~L. and {Craig}, M.~W. and {Dencheva}, N. and {Ginsburg}, A. and {VanderPlas}, J.~T. and {Bradley}, L.~D. and {P{\'e}rez-Su{\'a}rez}, D. and {de Val-Borro}, M. and {Aldcroft}, T.~L. and {Cruz}, K.~L. and {Robitaille}, T.~P. and {Tollerud}, E.~J. and {Ardelean}, C. and {Babej}, T. and {Bach}, Y.~P. and {Bachetti}, M. and {Bakanov}, A.~V. and {Bamford}, S.~P. and {Barentsen}, G. and {Barmby}, P. and {Baumbach}, A. and {Berry}, K.~L. and {Biscani}, F. and {Boquien}, M. and {Bostroem}, K.~A. and {Bouma}, L.~G. and {Brammer}, G.~B. and {Bray}, E.~M. and {Breytenbach}, H. and {Buddelmeijer}, H. and {Burke}, D.~J. and {Calderone}, G. and {Cano Rodr{\'\i}guez}, J.~L. and {Cara}, M. and {Cardoso}, J.~V.~M. and {Cheedella}, S. and {Copin}, Y. and {Corrales}, L. and {Crichton}, D. and {D'Avella}, D. and {Deil}, C. and {Depagne}, {\'E}. and {Dietrich}, J.~P. and {Donath}, A. and {Droettboom}, M. and {Earl}, N. and {Erben}, T. and {Fabbro}, S. and {Ferreira}, L.~A. and {Finethy}, T. and {Fox}, R.~T. and {Garrison}, L.~H. and {Gibbons}, S.~L.~J. and {Goldstein}, D.~A. and {Gommers}, R. and {Greco}, J.~P. and {Greenfield}, P. and {Groener}, A.~M. and {Grollier}, F. and {Hagen}, A. and {Hirst}, P. and {Homeier}, D. and {Horton}, A.~J. and {Hosseinzadeh}, G. and {Hu}, L. and {Hunkeler}, J.~S. and {Ivezi{\'c}}, {\v{Z}}. and {Jain}, A. and {Jenness}, T. and {Kanarek}, G. and {Kendrew}, S. and {Kern}, N.~S. and {Kerzendorf}, W.~E. and {Khvalko}, A. and {King}, J. and {Kirkby}, D. and {Kulkarni}, A.~M. and {Kumar}, A. and {Lee}, A. and {Lenz}, D. and {Littlefair}, S.~P. and {Ma}, Z. and {Macleod}, D.~M. and {Mastropietro}, M. and {McCully}, C. and {Montagnac}, S. and {Morris}, B.~M. and {Mueller}, M. and {Mumford}, S.~J. and {Muna}, D. and {Murphy}, N.~A. and {Nelson}, S. and {Nguyen}, G.~H. and {Ninan}, J.~P. and {N{\"o}the}, M. and {Ogaz}, S. and {Oh}, S. and {Parejko}, J.~K. and {Parley}, N. and {Pascual}, S. and {Patil}, R. and {Patil}, A.~A. and {Plunkett}, A.~L. and {Prochaska}, J.~X. and {Rastogi}, T. and {Reddy Janga}, V. and {Sabater}, J. and {Sakurikar}, P. and {Seifert}, M. and {Sherbert}, L.~E. and {Sherwood-Taylor}, H. and {Shih}, A.~Y. and {Sick}, J. and {Silbiger}, M.~T. and {Singanamalla}, S. and {Singer}, L.~P. and {Sladen}, P.~H. and {Sooley}, K.~A. and {Sornarajah}, S. and {Streicher}, O. and {Teuben}, P. and {Thomas}, S.~W. and {Tremblay}, G.~R. and {Turner}, J.~E.~H. and {Terr{\'o}n}, V. and {van Kerkwijk}, M.~H. and {de la Vega}, A. and {Watkins}, L.~L. and {Weaver}, B.~A. and {Whitmore}, J.~B. and {Woillez}, J. and {Zabalza}, V. and {Astropy Contributors}},
        title = "{The Astropy Project: Building an Open-science Project and Status of the v2.0 Core Package}",
      journal = {\aj},
         year = 2018,
        month = sep,
       volume = {156},
       number = {3},
          eid = {123},
        pages = {123},
          doi = {10.3847/1538-3881/aabc4f},
archivePrefix = {arXiv},
       eprint = {1801.02634},
 primaryClass = {astro-ph.IM},
       adsurl = {https://ui.adsabs.harvard.edu/abs/2018AJ....156..123A}
}

@ARTICLE{2013A&A...558A..33A,
       author = {{Astropy Collaboration} and {Robitaille}, Thomas P. and
         {Tollerud}, Erik J. and {Greenfield}, Perry and {Droettboom}, Michael and
         {Bray}, Erik and {Aldcroft}, Tom and {Davis}, Matt and
         {Ginsburg}, Adam and {Price-Whelan}, Adrian M. and
         {Kerzendorf}, Wolfgang E. and {Conley}, Alexander and {Crighton}, Neil and
         {Barbary}, Kyle and {Muna}, Demitri and {Ferguson}, Henry and
         {Grollier}, Fr{\'e}d{\'e}ric and {Parikh}, Madhura M. and
         {Nair}, Prasanth H. and {Unther}, Hans M. and {Deil}, Christoph and
         {Woillez}, Julien and {Conseil}, Simon and {Kramer}, Roban and
         {Turner}, James E.~H. and {Singer}, Leo and {Fox}, Ryan and
         {Weaver}, Benjamin A. and {Zabalza}, Victor and {Edwards}, Zachary I. and
         {Azalee Bostroem}, K. and {Burke}, D.~J. and {Casey}, Andrew R. and
         {Crawford}, Steven M. and {Dencheva}, Nadia and {Ely}, Justin and
         {Jenness}, Tim and {Labrie}, Kathleen and {Lim}, Pey Lian and
         {Pierfederici}, Francesco and {Pontzen}, Andrew and {Ptak}, Andy and
         {Refsdal}, Brian and {Servillat}, Mathieu and {Streicher}, Ole},
        title = "{Astropy: A community Python package for astronomy}",
      journal = {\aap},
         year = "2013",
        month = "Oct",
       volume = {558},
          eid = {A33},
        pages = {A33},
          doi = {10.1051/0004-6361/201322068},
archivePrefix = {arXiv},
       eprint = {1307.6212},
 primaryClass = {astro-ph.IM},
       adsurl = {https://ui.adsabs.harvard.edu/abs/2013A&A...558A..33A}
}

@article{sofue2020co,
  title={CO-to-H2 conversion and spectral column density in molecular clouds: the variability of the X CO factor},
  author={Sofue, Yoshiaki and Kohno, Mikito},
  journal={Monthly Notices of the Royal Astronomical Society},
  volume={497},
  number={2},
  pages={1851--1861},
  year={2020},
  publisher={Oxford University Press}
}

@article{kaiser1997self,
  title={A self-similar model for extragalactic radio sources},
  author={Kaiser, Christian R and Alexander, Paul},
  journal={Monthly Notices of the Royal Astronomical Society},
  volume={286},
  number={1},
  pages={215--222},
  year={1997},
  publisher={Blackwell Science Ltd Oxford, UK}
}

@article{gallo2005dark,
  title={A dark jet dominates the power output of the stellar black hole Cygnus X-1},
  author={Gallo, Elena and Fender, Rob and Kaiser, Christian and Russell, David and Morganti, Raffaella and Oosterloo, Tom and Heinz, Sebastian},
  journal={Nature},
  volume={436},
  number={7052},
  pages={819--821},
  year={2005},
  publisher={Nature Publishing Group UK London}
}

@article{tetarenko2018mapping,
  title={Mapping jet--ISM interactions in X-ray binaries with ALMA: a GRS 1915+ 105 case study},
  author={Tetarenko, Alexandra Jean and Freeman, P and Rosolowsky, EW and Miller-Jones, James C. A. and Sivakoff, Gregory Robert},
  journal={Monthly Notices of the Royal Astronomical Society},
  volume={475},
  number={1},
  pages={448--468},
  year={2018},
  publisher={Oxford University Press}
}

@article{tetarenko2020jet,
  title={Jet--ISM interactions near the microquasars GRS 1758- 258 and 1E 1740.7- 2942},
  author={Tetarenko, A. J. and Rosolowsky, E. W. and Miller-Jones, J. C. A. and Sivakoff, GR},
  journal={Monthly Notices of the Royal Astronomical Society},
  volume={497},
  number={3},
  pages={3504--3524},
  year={2020},
  publisher={Oxford University Press}
}

@article{leroy2021phangs,
  title={PHANGS--ALMA data processing and pipeline},
  author={Leroy, Adam K and Hughes, Annie and Liu, Daizhong and Pety, J{\'e}r{\^o}me and Rosolowsky, Erik and Saito, Toshiki and Schinnerer, Eva and Schruba, Andreas and Usero, Antonio and Faesi, Christopher M and others},
  journal={The Astrophysical Journal Supplement Series},
  volume={255},
  number={1},
  pages={19},
  year={2021},
  publisher={IOP Publishing}
}

@article{reid2019trigonometric,
  title={Trigonometric parallaxes of high-mass star-forming regions: our view of the Milky Way},
  author={Reid, MJ and Menten, KM and Brunthaler, A and Zheng, XW and Dame, TM and Xu, Y and Li, J and Sakai, N and Wu, Y and Immer, K and others},
  journal={The Astrophysical Journal},
  volume={885},
  number={2},
  pages={131},
  year={2019},
  publisher={IOP Publishing}
}

@article{wenger2018kinematic,
  title={Kinematic distances: a Monte Carlo method},
  author={Wenger, Trey V and Balser, Dana S and Anderson, LD and Bania, TM},
  journal={The Astrophysical Journal},
  volume={856},
  number={1},
  pages={52},
  year={2018},
  publisher={IOP Publishing}
}

@book{wilson2009tools,
  title={Tools of radio astronomy},
  author={Wilson, Thomas L and Rohlfs, Kristen and H{\"u}ttemeister, Susanne},
  volume={5},
  year={2009},
  publisher={Springer}
}

@article{magorrian1998demography,
  title={The demography of massive dark objects in galaxycenters},
  author={Magorrian, John and Tremaine, Scott and Richstone, Douglas and Bender, Ralf and Bower, Gary and Dressler, Alan and Faber, SM and Gebhardt, Karl and Green, Richard and Grillmair, Carl and others},
  journal={The Astronomical Journal},
  volume={115},
  number={6},
  pages={2285},
  year={1998},
  publisher={IOP Publishing}
}

@article{mcnamara2007heating,
  title={Heating hot atmospheres with active galactic nuclei},
  author={McNamara, BR and Nulsen, PEJ},
  journal={Annu. Rev. Astron. Astrophys.},
  volume={45},
  number={1},
  pages={117--175},
  year={2007},
  publisher={Annual Reviews}
}

@article{castor1975interstellar,
  title={Interstellar bubbles},
  author={Castor, John and McCray, Richard and Weaver, Robert},
  journal={Astrophysical Journal, vol. 200, Sept. 1, 1975, pt. 2, p. L107-L110.},
  volume={200},
  pages={L107--L110},
  year={1975}
}

@article{russell2007jet,
  title={The jet-powered optical nebula of Cygnus X--1},
  author={Russell, David M and Fender, RP and Gallo, E and Kaiser, CR},
  journal={Monthly Notices of the Royal Astronomical Society},
  volume={376},
  number={3},
  pages={1341--1349},
  year={2007},
  publisher={Blackwell Publishing Ltd Oxford, UK}
}

@article{sell2015shell,
  title={Shell-shocked: the interstellar medium near Cygnus X-1},
  author={Sell, PH and Heinz, S and Richards, E and Maccarone, TJ and Russell, DM and Gallo, E and Fender, R and Markoff, S and Nowak, M},
  journal={Monthly Notices of the Royal Astronomical Society},
  volume={446},
  number={4},
  pages={3579--3592},
  year={2015},
  publisher={Oxford University Press}
}

@article{kaiser2004revision,
  title={Revision of the properties of the GRS 1915+ 105 jets: clues from the large-scale structure},
  author={Kaiser, Christian R and Gunn, Katherine F and Brocksopp, Catherine and Sokoloski, Jennifer L},
  journal={The Astrophysical Journal},
  volume={612},
  number={1},
  pages={332},
  year={2004},
  publisher={IOP Publishing}
}

@article{blandford1982hydromagnetic,
  title={Hydromagnetic flows from accretion discs and the production of radio jets},
  author={Blandford, Roger D and Payne, DG},
  journal={Monthly Notices of the Royal Astronomical Society},
  volume={199},
  number={4},
  pages={883--903},
  year={1982},
  publisher={Oxford University Press Oxford, UK}
}

@article{blandford1977electromagnetic,
  title={Electromagnetic extraction of energy from Kerr black holes},
  author={Blandford, Roger D and Znajek, Roman L},
  journal={Monthly Notices of the Royal Astronomical Society},
  volume={179},
  number={3},
  pages={433--456},
  year={1977},
  publisher={Oxford University Press Oxford, UK}
}

@article{motta2025meerkat,
  title={MeerKAT discovers a jet-driven bow shock near GRS 1915+ 105-How an invisible large-scale jet sculpts a microquasar’s environment},
  author={Motta, SE and Atri, P and Matthews, James H and van den Eijnden, Jakob and Fender, Rob P and Miller-Jones, James C. A. and Heywood, Ian and Woudt, Patrick},
  journal={Astronomy \& Astrophysics},
  volume={696},
  pages={A222},
  year={2025},
  publisher={EDP Sciences}
}

@article{van2007computer,
  title={A computer program for fast non-LTE analysis of interstellar line spectra-With diagnostic plots to interpret observed line intensity ratios},
  author={Van der Tak, FFS and Black, John H and Sch{\"o}ier, FL and Jansen, DJ and van Dishoeck, Ewine F},
  journal={Astronomy \& Astrophysics},
  volume={468},
  number={2},
  pages={627--635},
  year={2007},
  publisher={EDP Sciences}
}

@misc{2019zndo...2573901G,
       author = {{Ginsburg}, Adam and {Koch}, Eric and {Robitaille}, Thomas and {Beaumont}, Chris and {Adamginsburg} and {ZuHone}, John and {Sipocz}, Brigitta and {Patra}, Sushobhana and {Jones}, Craig and {Lim}, P.~L. and {Rosolowsky}, Erik and {Stern}, Kris and {Earl}, Nicholas and {De Val-Borro}, Miguel and {Jrobbfed} and {Shuokong} and {Kepley}, Amanda and {Sokolov}, Vlas and {Badger}, The Gitter and {Maret}, S{\'e}bastien and {Garrido}, Juli{\'a}n and {Booker}, Joseph and {Tollerud}, Erik},
        title = "{radio-astro-tools/spectral-cube: v0.4.4}",
         year = 2019,
        month = feb,
          eid = {10.5281/zenodo.2573901},
          doi = {10.5281/zenodo.2573901},
      version = {v0.4.4},
    publisher = {Zenodo},
       adsurl = {https://ui.adsabs.harvard.edu/abs/2019zndo...2573901G}
}

@misc{2016ascl.soft08010G,
       author = {{Ginsburg}, Adam and {Robitaille}, Thomas and {Beaumont}, Chris},
        title = "{pvextractor: Position-Velocity Diagram Extractor}",
 howpublished = {Astrophysics Source Code Library, record ascl:1608.010},
         year = 2016,
        month = aug,
          eid = {ascl:1608.010},
       adsurl = {https://ui.adsabs.harvard.edu/abs/2016ascl.soft08010G}
}

@article{bosch2026constraining,
  title={Constraining the Jet Energetics of the Transient X-Ray Binaries MAXI J1348- 630 and MAXI J1820+ 070 through Calorimetry},
  author={Bosch-Cabot, Pau and Tetarenko, Alexandra J. and Rosolowsky, Erik and Carotenuto, Francesco and Miller-Jones, James C. A. and Russell, David M and Corbel, St{\'e}phane and Russell, Thomas D and Sivakoff, Gregory R},
  journal={The Astrophysical Journal},
  volume={997},
  number={1},
  pages={64},
  year={2026},
  publisher={IOP Publishing}
}

@ARTICLE{gies2008stellar,
       author = {{Gies}, D.~R. and {Bolton}, C.~T. and {Blake}, R.~M. and {Caballero-Nieves}, S.~M. and {Crenshaw}, D.~M. and {Hadrava}, P. and {Herrero}, A. and {Hillwig}, T.~C. and {Howell}, S.~B. and {Huang}, W. and {Kaper}, L. and {Koubsk{\'y}}, P. and {McSwain}, M.~V.},
        title = "{Stellar Wind Variations during the X-Ray High and Low States of Cygnus X-1}",
      journal = {\apj},
         year = 2008,
        month = may,
       volume = {678},
       number = {2},
        pages = {1237-1247},
          doi = {10.1086/586690},
archivePrefix = {arXiv},
       eprint = {0801.4286},
 primaryClass = {astro-ph},
       adsurl = {https://ui.adsabs.harvard.edu/abs/2008ApJ...678.1237G}
}

@article{herrero1995fundamental,
  title={Fundamental parameters of galactic luminous OB stars. II. A spectroscopic analysis of HDE 226868 and the mass of Cygnus X-1.},
  author={Herrero, A and Kudritzki, RP and Gabler, R and Vilchez, JM and Gabler, A},
  journal={Astronomy and Astrophysics, v. 297, p. 556},
  volume={297},
  pages={556},
  year={1995}
}

@article{gies2003wind,
  title={Wind accretion and state transitions in Cygnus X-1},
  author={Gies, DR and Bolton, CT and Thomson, JR and Huang, W and McSwain, MV and Riddle, RL and Wang, Z and Wiita, PJ and Wingert, DW and Cs{\'a}k, B and others},
  journal={The Astrophysical Journal},
  volume={583},
  number={1},
  pages={424},
  year={2003},
  publisher={IOP Publishing}
}

@article{miller2021cygnus,
  title={Cygnus X-1 contains a 21--solar mass black hole—Implications for massive star winds},
  author={Miller-Jones, James C. A. and Bahramian, Arash and Orosz, Jerome A and Mandel, Ilya and Gou, Lijun and Maccarone, Thomas J and Neijssel, Coenraad J and Zhao, Xueshan and Zi{\'o}{\l}kowski, Janusz and Reid, Mark J and others},
  journal={Science},
  volume={371},
  number={6533},
  pages={1046--1049},
  year={2021},
  publisher={American Association for the Advancement of Science}
}

@article{carretero2025observational,
  title={An observational study of rotation and binarity of Galactic O-type runaway stars},
  author={Carretero-Castrillo, M and Rib{\'o}, M and Paredes, JM and Holgado, G and Mart{\'\i}nez-Sebasti{\'a}n, C and Sim{\'o}n-D{\'\i}az, S},
  journal={arXiv preprint arXiv:2510.21577},
  year={2025}
}

@article{atri2025quantifying,
  title={Quantifying jet--interstellar medium interactions in Cyg X-1: Insights from dual-frequency bow shock detection with MeerKAT},
  author={Atri, P and Motta, SE and van den Eijnden, Jakob and Matthews, James H and Miller-Jones, James CA and Fender, Rob and Williams-Baldwin, David and Heywood, Ian and Woudt, Patrick},
  journal={Astronomy \& Astrophysics},
  volume={696},
  pages={A223},
  year={2025},
  publisher={EDP Sciences}
}

@article{lai2024characterisation,
  title={Characterisation of the stellar wind in Cyg X-1 via modelling of colour-colour diagrams},
  author={Lai, Eleonora Veronica and De Marco, B{\'a}rbara and Cavecchi, Yuri and El Mellah, Ileyk and Cinus, M and Diez, CM and Grinberg, Victoria and Zdziarski, Andrzej A and Uttley, Phil and Bachetti, Matteo and others},
  journal={Astronomy \& Astrophysics},
  volume={691},
  pages={A78},
  year={2024},
  publisher={EDP Sciences}
}

@article{ramachandran2025comprehensive,
  title={Comprehensive UV and optical spectral analysis of Cygnus X-1-Stellar and wind parameters, abundances, and evolutionary implications},
  author={Ramachandran, V and Sander, AAC and Oskinova, LM and Sch{\"o}sser, EC and Pauli, D and Hamann, W-R and Mahy, L and Bernini-Peron, M and Brigitte, M and Kub{\'a}tov{\'a}, B},
  journal={Astronomy \& Astrophysics},
  volume={698},
  pages={A37},
  year={2025},
  publisher={EDP Sciences}
}

@article{komugi2025alma,
  title={ALMA FACTS. II. Large Scale Variations in the 12CO (J= 2--1) to 12CO (J= 1--0) Line Ratio in Nearby Galaxies},
  author={Komugi, Shinya and Sawada, Tsuyoshi and Koda, Jin and Egusa, Fumi and Maeda, Fumiya and Hirota, Akihiko and Lee, Amanda M},
  journal={The Astrophysical Journal},
  volume={980},
  number={1},
  pages={126},
  year={2025},
  publisher={IOP Publishing}
}

@article{puls2008mass,
  title={Mass loss from hot massive stars},
  author={Puls, Joachim and Vink, Jorick S and Najarro, Francisco},
  journal={The Astronomy and Astrophysics Review},
  volume={16},
  number={3},
  pages={209--325},
  year={2008},
  publisher={Springer}
}

@article{van2009dense,
  title={Dense and warm molecular gas in the envelopes and outflows of southern low-mass protostars},
  author={van Kempen, TA and van Dishoeck, EF and Hogerheijde, MR and G{\"u}sten, R},
  journal={Astronomy \& Astrophysics},
  volume={508},
  number={1},
  pages={259--274},
  year={2009},
  publisher={EDP Sciences}
}

@article{nesvadba2010energetics,
  title={Energetics of the molecular gas in the H2 luminous radio galaxy 3C 326: Evidence for negative AGN feedback},
  author={Nesvadba, NPH and Boulanger, Francois and Salom{\'e}, Philippe and Guillard, P and Lehnert, Matthew D and Ogle, P and Appleton, P and Falgarone, Edith and Des Forets, G Pineau},
  journal={Astronomy \& Astrophysics},
  volume={521},
  pages={A65},
  year={2010},
  publisher={EDP Sciences}
}

@article{krause2023jet,
  title={Jet feedback in star-forming galaxies},
  author={Krause, Martin GH},
  journal={Galaxies},
  volume={11},
  number={1},
  pages={29},
  year={2023},
  publisher={MDPI}
}

@article{mariani2025meerkat,
  title={A MeerKAT view of the parsec-scale jets in the black-hole X-ray binary GRS 1758--258},
  author={Mariani, I and Motta, SE and Atri, P and Matthews, JH and Fender, RP and Mart{\'\i}, J and Luque-Escamilla, PL and Heywood, I},
  journal={Astronomy \& Astrophysics},
  volume={704},
  pages={A239},
  year={2025},
  publisher={EDP Sciences}
}

@article{heinz2008blazing,
  title={Blazing trails: microquasars as head-tail sources and the seeding of magnetized plasma into the ISM},
  author={Heinz, S and Grimm, HJ and Sunyaev, RA and Fender, RP},
  journal={The Astrophysical Journal},
  volume={686},
  number={2},
  pages={1145},
  year={2008},
  publisher={IOP Publishing}
}

@article{alfaro2024ultra,
  title={Ultra-high-energy gamma-ray bubble around microquasar V4641 Sgr},
  author={Alfaro, R and Alvarez, C and Arteaga-Vel{\'a}zquez, JC and Avila Rojas, D and Ayala Solares, HA and Babu, R and Belmont-Moreno, E and Caballero-Mora, KS and Capistr{\'a}n, T and Carrami{\~n}ana, A and others},
  journal={Nature},
  volume={634},
  number={8034},
  pages={557--560},
  year={2024},
  publisher={Nature Publishing Group UK London}
}

@article{churazov2002cooling,
  title={Cooling flows as a calorimeter of active galactic nucleus mechanical power},
  author={Churazov, Eugene and Sunyaev, R and Forman, W and B{\"o}hringer, H},
  journal={Monthly Notices of the Royal Astronomical Society},
  volume={332},
  number={3},
  pages={729--734},
  year={2002},
  publisher={The Royal Astronomical Society}
}

@article{fender2006transient,
  title={A transient relativistic radio jet from Cygnus X-1},
  author={Fender, Rob P and Stirling, AM and Spencer, RE and Brown, I and Pooley, GG and Muxlow, TWB and Miller-Jones, J. C. A.},
  journal={Monthly Notices of the Royal Astronomical Society},
  volume={369},
  number={2},
  pages={603--607},
  year={2006},
  publisher={Blackwell Publishing Ltd Oxford, UK}
}

@article{sugimoto2017orbital,
  title={Orbital modulations of X-ray light curves of Cygnus X-1 in its low/hard and high/soft states},
  author={Sugimoto, Juri and Kitamoto, Shunji and Mihara, Tatehiro and Matsuoka, Masaru},
  journal={Publications of the Astronomical Society of Japan},
  volume={69},
  number={3},
  pages={52},
  year={2017},
  publisher={Oxford University Press}
}

@article{trigo2021search,
  title={A search for signatures of interactions of X-ray binary outflows with their environments with ALMA},
  author={D{\'\i}az-Trigo, M  and Petry, D and Humphreys, E and Impellizzeri, CMV and Liu, Hauyu Baobab},
  journal={Astronomy \& Astrophysics},
  volume={650},
  pages={A37},
  year={2021},
  publisher={EDP Sciences}
}

@article{bosch2011termination,
  title={The termination region of high-mass microquasar jets},
  author={Bosch-Ramon, V and Perucho, M and Bordas, P},
  journal={Astronomy \& Astrophysics},
  volume={528},
  pages={A89},
  year={2011},
  publisher={EDP Sciences}
}

@article{comeron1998numerical,
  title={Numerical simulations of wind bow shocks produced by runaway OB stars},
  author={Comer{\'o}n, F and Kaper, Lex},
  journal={Astronomy and Astrophysics, v. 338, p. 273-291 (1998)},
  volume={338},
  pages={273--291},
  year={1998}
}

@article{martinez2023probing,
  title={Probing the non-thermal physics of stellar bow shocks using radio observations},
  author={Martinez, Javier Rodrigo and del Palacio, Santiago and Bosch-Ramon, Valent{\'\i}},
  journal={Astronomy \& Astrophysics},
  volume={680},
  pages={A99},
  year={2023},
  publisher={EDP Sciences}
}

@article{brocksopp1998improved,
  title={An improved orbital ephemeris for Cygnus X-1},
  author={Brocksopp, C and Tarasov, AE and Lyuty, VM and Roche, P},
  journal={arXiv preprint astro-ph/9812077},
  year={1998}
}

@article{hirai2021conditions,
  title={Conditions for accretion disc formation and observability of wind-accreting X-ray binaries},
  author={Hirai, Ryosuke and Mandel, Ilya},
  journal={Publications of the Astronomical Society of Australia},
  volume={38},
  pages={e056},
  year={2021},
  publisher={Cambridge University Press}
}

@ARTICLE{nagarajan2025mixed,
       author = {{Nagarajan}, Pranav and {El-Badry}, Kareem},
        title = "{Mixed Origins: Strong Natal Kicks for Some Black Holes and None for Others}",
      journal = {\pasp},
         year = 2025,
        month = mar,
       volume = {137},
       number = {3},
          eid = {034203},
        pages = {034203},
          doi = {10.1088/1538-3873/adb6d6},
archivePrefix = {arXiv},
       eprint = {2411.16847},
 primaryClass = {astro-ph.GA},
       adsurl = {https://ui.adsabs.harvard.edu/abs/2025PASP..137c4203N}
}

@article{grinberg2013long,
  title={Long term variability of Cygnus X-1-V. State definitions with all sky monitors},
  author={Grinberg, V and Hell, N and Pottschmidt, Katja and B{\"o}ck, M and Nowak, MA and Rodriguez, J and Bodaghee, A and Bel, M Cadolle and Case, GL and Hanke, M and others},
  journal={Astronomy \& Astrophysics},
  volume={554},
  pages={A88},
  year={2013},
  publisher={EDP Sciences}
}

@article{mirabel2003formation,
  title={Formation of a black hole in the dark},
  author={Mirabel, I F{\'e}lix and Rodrigues, Irapuan},
  journal={Science},
  volume={300},
  number={5622},
  pages={1119--1120},
  year={2003},
  publisher={American Association for the Advancement of Science}
}

@ARTICLE{zealey1980interaction,
       author = {{Zealey}, W.~J. and {Dopita}, M.~A. and {Malin}, D.~F.},
        title = "{The interaction between the relativistic jets of SS433 and the interstellar medium}",
      journal = {\mnras},
         year = 1980,
        month = sep,
       volume = {192},
        pages = {731-743},
          doi = {10.1093/mnras/192.4.731},
       adsurl = {https://ui.adsabs.harvard.edu/abs/1980MNRAS.192..731Z}
}

@ARTICLE{mcleod2019optical,
       author = {{McLeod}, A.~F. and {Scaringi}, S. and {Soria}, R. and {Pakull}, M.~W. and {Urquhart}, R. and {Maccarone}, T.~J. and {Knigge}, C. and {Miller-Jones}, J.~C.~A. and {Plotkin}, R.~M. and {Motch}, C. and {Kruijssen}, J.~M.~D. and {Schruba}, A.},
        title = "{Optical IFU spectroscopy of a bipolar microquasar jet in NGC 300}",
      journal = {\mnras},
         year = 2019,
        month = may,
       volume = {485},
       number = {3},
        pages = {3476-3485},
          doi = {10.1093/mnras/stz614},
archivePrefix = {arXiv},
       eprint = {1903.00005},
 primaryClass = {astro-ph.HE},
       adsurl = {https://ui.adsabs.harvard.edu/abs/2019MNRAS.485.3476M}
}

@ARTICLE{hess2024acceleration,
       author = {{H.~E.~S.~S. Collaboration} and {Aharonian}, F. and {Ait Benkhali}, F. and {Aschersleben}, J. and {Ashkar}, H. and {Backes}, M. and {Barbosa Martins}, V. and {Batzofin}, R. and {Becherini}, Y. and {Berge}, D. and {Bernl{\"o}hr}, K. and {Bi}, B. and {B{\"o}ttcher}, M. and {Boisson}, C. and {Bolmont}, J. and {de Lavergne}, M. de Bony and {Borowska}, J. and {Bouyahiaoui}, M. and {Breuhaus}, M. and {Brose}, R. and {Brown}, A.~M. and {Brun}, F. and {Bruno}, B. and {Bulik}, T. and {Burger-Scheidlin}, C. and {Caroff}, S. and {Casanova}, S. and {Cecil}, R. and {Celic}, J. and {Cerruti}, M. and {Chand}, T. and {Chandra}, S. and {Chen}, A. and {Chibueze}, J. and {Chibueze}, O. and {Cotter}, G. and {Dai}, S. and {Mbarubucyeye}, J. Damascene and {Djannati-Ata{\"\i}}, A. and {Dmytriiev}, A. and {Doroshenko}, V. and {Egberts}, K. and {Einecke}, S. and {Ernenwein}, J.-P. and {Filipovic}, M. and {Fontaine}, G. and {F{\"u}{\ss}ling}, M. and {Funk}, S. and {Gabici}, S. and {Ghafourizadeh}, S. and {Giavitto}, G. and {Glawion}, D. and {Glicenstein}, J.-F. and {Grolleron}, G. and {Haerer}, L. and {Hinton}, J.~A. and {Hofmann}, W. and {Holch}, T.~L. and {Holler}, M. and {Horns}, D. and {Jamrozy}, M. and {Jankowsky}, F. and {Jardin-Blicq}, A. and {Joshi}, V. and {Jung-Richardt}, I. and {Kasai}, E. and {Katarzy{\'n}ski}, K. and {Khatoon}, R. and {Kh{\'e}lifi}, B. and {Klepser}, S. and {Klu{\'z}niak}, W. and {Komin}, Nu. and {Kosack}, K. and {Kostunin}, D. and {Kundu}, A. and {Lang}, R.~G. and {Le Stum}, S. and {Leitl}, F. and {Lemi{\`e}re}, A. and {Lenain}, J.-P. and {Leuschner}, F. and {Lohse}, T. and {Luashvili}, A. and {Lypova}, I. and {Mackey}, J. and {Malyshev}, D. and {Malyshev}, D. and {Marandon}, V. and {Marchegiani}, P. and {Marcowith}, A. and {Mart{\'\i}-Devesa}, G. and {Marx}, R. and {Mehta}, A. and {Mitchell}, A. and {Moderski}, R. and {Mohrmann}, L. and {Montanari}, A. and {Moulin}, E. and {Murach}, T. and {Nakashima}, K. and {de Naurois}, M. and {Niemiec}, J. and {Noel}, A. Priyana and {Ohm}, S. and {Olivera-Nieto}, L. and {de Ona Wilhelmi}, E. and {Ostrowski}, M. and {Panny}, S. and {Panter}, M. and {Parsons}, R.~D. and {Peron}, G. and {Prokhorov}, D.~A. and {P{\"u}hlhofer}, G. and {Punch}, M. and {Quirrenbach}, A. and {Reichherzer}, P. and {Reimer}, A. and {Reimer}, O. and {Ren}, H. and {Renaud}, M. and {Reville}, B. and {Rieger}, F. and {Rowell}, G. and {Rudak}, B. and {Ricarte}, H. Rueda and {Ruiz-Velasco}, E. and {Sahakian}, V. and {Salzmann}, H. and {Santangelo}, A. and {Sasaki}, M. and {Sch{\"a}fer}, J. and {Sch{\"u}ssler}, F. and {Schwanke}, U. and {Shapopi}, J.~N.~S. and {Sol}, H. and {Specovius}, A. and {Spencer}, S. and {Stawarz}, L. and {Steenkamp}, R. and {Steinmassl}, S. and {Steppa}, C. and {Streil}, K. and {Sushch}, I. and {Suzuki}, H. and {Takahashi}, T. and {Tanaka}, T. and {Taylor}, A.~M. and {Terrier}, R. and {Tsirou}, M. and {Tsuji}, N. and {Unbehaun}, T. and {van Eldik}, C. and {Vecchi}, M. and {Veh}, J. and {Venter}, C. and {Vink}, J. and {Wach}, T. and {Wagner}, S.~J. and {Werner}, F. and {White}, R. and {Wierzcholska}, A. and {Wong}, Yu Wun and {Zacharias}, M. and {Zargaryan}, D. and {Zdziarski}, A.~A. and {Zech}, A. and {Zouari}, S. and {{\.Z}ywucka}, N.},
        title = "{Acceleration and transport of relativistic electrons in the jets of the microquasar SS 433}",
      journal = {Science},
         year = 2024,
        month = jan,
       volume = {383},
       number = {6681},
        pages = {402-406},
          doi = {10.1126/science.adi2048},
archivePrefix = {arXiv},
       eprint = {2401.16019},
 primaryClass = {astro-ph.HE},
       adsurl = {https://ui.adsabs.harvard.edu/abs/2024Sci...383..402H}
}

@ARTICLE{brigitte2025disentangling,
       author = {{Brigitte}, M. and {Hadrava}, P. and {Kub{\'a}tov{\'a}}, B. and {Cabezas}, M. and {Svoboda}, J. and {{\v{S}}lechta}, M. and {Skarka}, M. and {Alabarta}, K. and {Maryeva}, O. and {Russell}, D.~M. and {Baglio}, M.~C.},
        title = "{Disentangling the stellar atmosphere and the focused wind in different accretion states of Cygnus X-1}",
      journal = {\aap},
         year = 2025,
        month = mar,
       volume = {695},
          eid = {A115},
        pages = {A115},
          doi = {10.1051/0004-6361/202450910},
archivePrefix = {arXiv},
       eprint = {2502.07989},
 primaryClass = {astro-ph.HE},
       adsurl = {https://ui.adsabs.harvard.edu/abs/2025A&A...695A.115B}
}

@ARTICLE{munoz2016regulation,
       author = {{Mu{\~n}oz-Darias}, T. and {Casares}, J. and {Mata S{\'a}nchez}, D. and {Fender}, R.~P. and {Armas Padilla}, M. and {Linares}, M. and {Ponti}, G. and {Charles}, P.~A. and {Mooley}, K.~P. and {Rodriguez}, J.},
        title = "{Regulation of black-hole accretion by a disk wind during a violent outburst of V404 Cygni}",
      journal = {\nat},
         year = 2016,
        month = jun,
       volume = {534},
       number = {7605},
        pages = {75-78},
          doi = {10.1038/nature17446},
archivePrefix = {arXiv},
       eprint = {1605.02358},
 primaryClass = {astro-ph.HE},
       adsurl = {https://ui.adsabs.harvard.edu/abs/2016Natur.534...75M}
}

@ARTICLE{schonrich2010local,
       author = {{Sch{\"o}nrich}, Ralph and {Binney}, James and {Dehnen}, Walter},
        title = "{Local kinematics and the local standard of rest}",
      journal = {\mnras},
         year = 2010,
        month = apr,
       volume = {403},
       number = {4},
        pages = {1829-1833},
          doi = {10.1111/j.1365-2966.2010.16253.x},
archivePrefix = {arXiv},
       eprint = {0912.3693},
 primaryClass = {astro-ph.GA},
       adsurl = {https://ui.adsabs.harvard.edu/abs/2010MNRAS.403.1829S}
}

@ARTICLE{martinez2022nonthermal,
       author = {{Martinez}, J.~R. and {del Palacio}, S. and {Bosch-Ramon}, V. and {Romero}, G.~E.},
        title = "{Non-thermal emission in hyper-velocity and semi-relativistic stars}",
      journal = {\aap},
         year = 2022,
        month = may,
       volume = {661},
          eid = {A102},
        pages = {A102},
          doi = {10.1051/0004-6361/202142727},
archivePrefix = {arXiv},
       eprint = {2203.09686},
 primaryClass = {astro-ph.HE},
       adsurl = {https://ui.adsabs.harvard.edu/abs/2022A&A...661A.102M}
}

@ARTICLE{marti1996search,
       author = {{Marti}, J. and {Rodriguez}, L.~F. and {Mirabel}, I.~F. and {Paredes}, J.~M.},
        title = "{A search for arcminute-scale radio jets in Cygnus X-1.}",
      journal = {\aap},
         year = 1996,
        month = feb,
       volume = {306},
        pages = {449},
       adsurl = {https://ui.adsabs.harvard.edu/abs/1996A&A...306..449M}
}

@ARTICLE{prabu2026jet,
       author = {{Prabu}, S. and {Miller-Jones}, J.~C.~A. and {Bahramian}, A. and {Bosch-Ramon}, V. and {Heinz}, S. and {Tingay}, S.~J. and {Wood}, C.~M. and {Tetarenko}, A.~J. and {O'Doherty}, T.~N. and {Tudose}, V.},
        title = "{A jet bent by a stellar wind in the black hole X-ray binary Cygnus X-1}",
      journal = {Nature Astronomy},
         year = 2026,
        month = apr,
          doi = {10.1038/s41550-026-02828-3},
       adsurl = {https://ui.adsabs.harvard.edu/abs/2026NatAs.tmp...83P}
}

@article{marti2017galactic,
  title={A galactic microquasar mimicking winged radio galaxies},
  author={Mart{\'\i}, Josep and Luque-Escamilla, Pedro L and Bosch-Ramon, Valent{\'\i} and Paredes, Josep M},
  journal={Nature Communications},
  volume={8},
  number={1},
  pages={1757},
  year={2017},
  publisher={Nature Publishing Group UK London}
}

@ARTICLE{2010AJ....140.1868W,
       author = {{Wright}, Edward L. and {Eisenhardt}, Peter R.~M. and {Mainzer}, Amy K. and {Ressler}, Michael E. and {Cutri}, Roc M. and {Jarrett}, Thomas and {Kirkpatrick}, J. Davy and {Padgett}, Deborah and {McMillan}, Robert S. and {Skrutskie}, Michael and {Stanford}, S.~A. and {Cohen}, Martin and {Walker}, Russell G. and {Mather}, John C. and {Leisawitz}, David and {Gautier}, III, Thomas N. and {McLean}, Ian and {Benford}, Dominic and {Lonsdale}, Carol J. and {Blain}, Andrew and {Mendez}, Bryan and {Irace}, William R. and {Duval}, Valerie and {Liu}, Fengchuan and {Royer}, Don and {Heinrichsen}, Ingolf and {Howard}, Joan and {Shannon}, Mark and {Kendall}, Martha and {Walsh}, Amy L. and {Larsen}, Mark and {Cardon}, Joel G. and {Schick}, Scott and {Schwalm}, Mark and {Abid}, Mohamed and {Fabinsky}, Beth and {Naes}, Larry and {Tsai}, Chao-Wei},
        title = "{The Wide-field Infrared Survey Explorer (WISE): Mission Description and Initial On-orbit Performance}",
      journal = {\aj},
         year = 2010,
        month = dec,
       volume = {140},
       number = {6},
        pages = {1868-1881},
          doi = {10.1088/0004-6256/140/6/1868},
archivePrefix = {arXiv},
       eprint = {1008.0031},
 primaryClass = {astro-ph.IM},
       adsurl = {https://ui.adsabs.harvard.edu/abs/2010AJ....140.1868W}
}

@misc{CARTA,
  doi = {10.5281/ZENODO.3377984},
  url = {https://zenodo.org/doi/10.5281/zenodo.3377984},
  author = {Angus Comrie   and Kuo-Song Wang   and Yu-Hsuan Hwang   and Adrianna Pińska   and Pamela Harris   and Hou,  Kuan-Chou and Gao,  Zhen-Kai and Aikema,  David and Huang,  Po-Sheng and Sokolowski,  Marcin and van Zyl,  Michaela and Rob Simmonds  },
  title = {CARTA: The Cube Analysis and Rendering Tool for Astronomy},
  publisher = {Zenodo},
  year = {2026},
  copyright = {Creative Commons Attribution 4.0 International}
}

@misc{gildas,
       author = {{Gildas Team}},
        title = "{GILDAS: Grenoble Image and Line Data Analysis Software}",
 howpublished = {Astrophysics Source Code Library, record ascl:1305.010},
         year = 2013,
        month = may,
          eid = {ascl:1305.010},
archivePrefix = {ascl},
       eprint = {1305.010},
       adsurl = {https://ui.adsabs.harvard.edu/abs/2013ascl.soft05010G}
}

@misc{irsa1,
  doi = {10.26131/IRSA1},
  url = {https://catcopy.ipac.caltech.edu/dois/doi.php?id=10.26131/IRSA1},
  author = {
  Edward L. Wright and
  Peter R. M. Eisenhardt and
  Amy K. Mainzer and
  Michael E. Ressler and
  Roc M. Cutri and
  Thomas Jarrett and
  J. Davy Kirkpatrick and
  Deborah Padgett and
  Robert S. McMillan and
  Michael Skrutskie and
  S. A. Stanford and
  Martin Cohen and
  Russell G. Walker and
  John C. Mather and
  David Leisawitz and
  Thomas N. Gautier III and
  Ian McLean and
  Dominic Benford and
  Carol J. Lonsdale and
  Andrew Blain and
  Bryan Mendez and
  William R. Irace and
  Valerie Duval and
  Fengchuan Liu and
  Don Royer and
  Ingolf Heinrichsen and
  Joan Howard and
  Mark Shannon and
  Martha Kendall and
  Amy L. Walsh and
  Mark Larsen and
  Joel G. Cardon and
  Scott Schick and
  Mark Schwalm and
  Mohamed Abid and
  Beth Fabinsky and
  Larry Naes and
  ChaoWei Tsai
},
  title = {AllWISE Source Catalog},
  publisher = {IPAC},
  year = {2019},
  copyright = {Creative Commons Attribution 4.0 International}
}

@BOOK{dopita2003astrophysics,
       author = {{Dopita}, Michael A. and {Sutherland}, Ralph S.},
        title = "{Astrophysics of the diffuse universe}",
         year = 2003,
          doi = {10.1007/978-3-662-05866-4},
          publisher={Springer},
       adsurl = {https://ui.adsabs.harvard.edu/abs/2003adu..book.....D}
}

@ARTICLE{meyer2014models,
       author = {{Meyer}, D.~M.-A. and {Mackey}, J. and {Langer}, N. and {Gvaramadze}, V.~V. and {Mignone}, A. and {Izzard}, R.~G. and {Kaper}, L.},
        title = "{Models of the circumstellar medium of evolving, massive runaway stars moving through the Galactic plane}",
      journal = {\mnras},
         year = 2014,
        month = nov,
       volume = {444},
       number = {3},
        pages = {2754-2775},
          doi = {10.1093/mnras/stu1629},
archivePrefix = {arXiv},
       eprint = {1408.2828},
 primaryClass = {astro-ph.SR},
       adsurl = {https://ui.adsabs.harvard.edu/abs/2014MNRAS.444.2754M}
}

@ARTICLE{acreman2016modelling,
       author = {{Acreman}, David M. and {Stevens}, Ian R. and {Harries}, Tim J.},
        title = "{Modelling multiwavelength observational characteristics of bow shocks from runaway early-type stars}",
      journal = {\mnras},
         year = 2016,
        month = feb,
       volume = {456},
       number = {1},
        pages = {136-145},
          doi = {10.1093/mnras/stv2632},
archivePrefix = {arXiv},
       eprint = {1511.03059},
 primaryClass = {astro-ph.SR},
       adsurl = {https://ui.adsabs.harvard.edu/abs/2016MNRAS.456..136A}
}

@ARTICLE{sutherland1993cooling,
       author = {{Sutherland}, Ralph S. and {Dopita}, M.~A.},
        title = "{Cooling Functions for Low-Density Astrophysical Plasmas}",
      journal = {\apjs},
         year = 1993,
        month = sep,
       volume = {88},
        pages = {253},
          doi = {10.1086/191823},
       adsurl = {https://ui.adsabs.harvard.edu/abs/1993ApJS...88..253S}
}
\bibliographystyle{aasjournalv7}

\appendix

\section{Infrared observations of the Cygnus X--1 environment} \label{sec:IR}

\begin{figure*}[h!]
    \centering
    \includegraphics[width=0.48\textwidth]{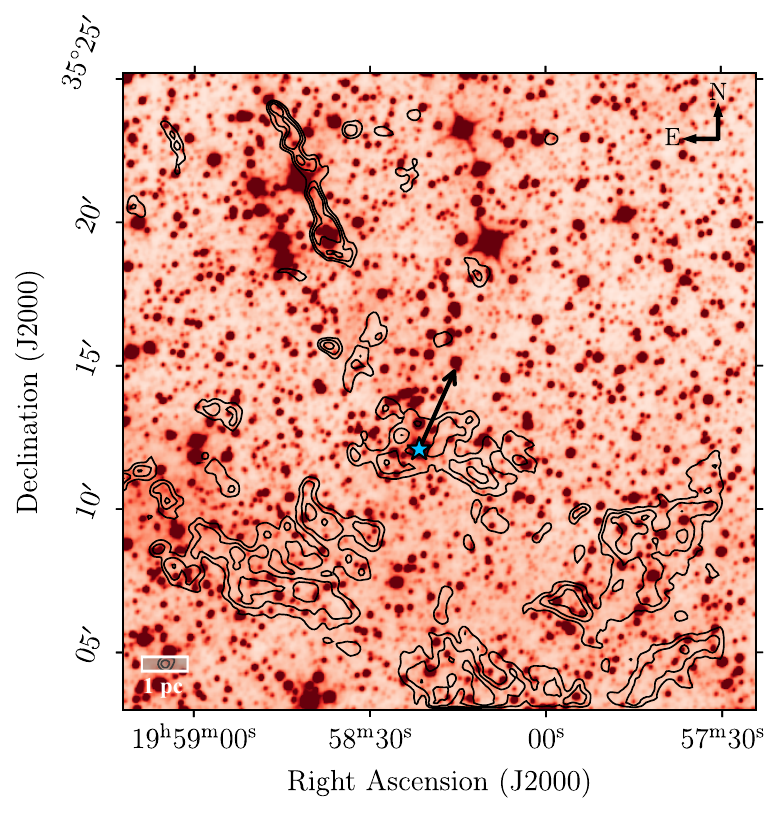}\hfill
    \includegraphics[width=0.48\textwidth]{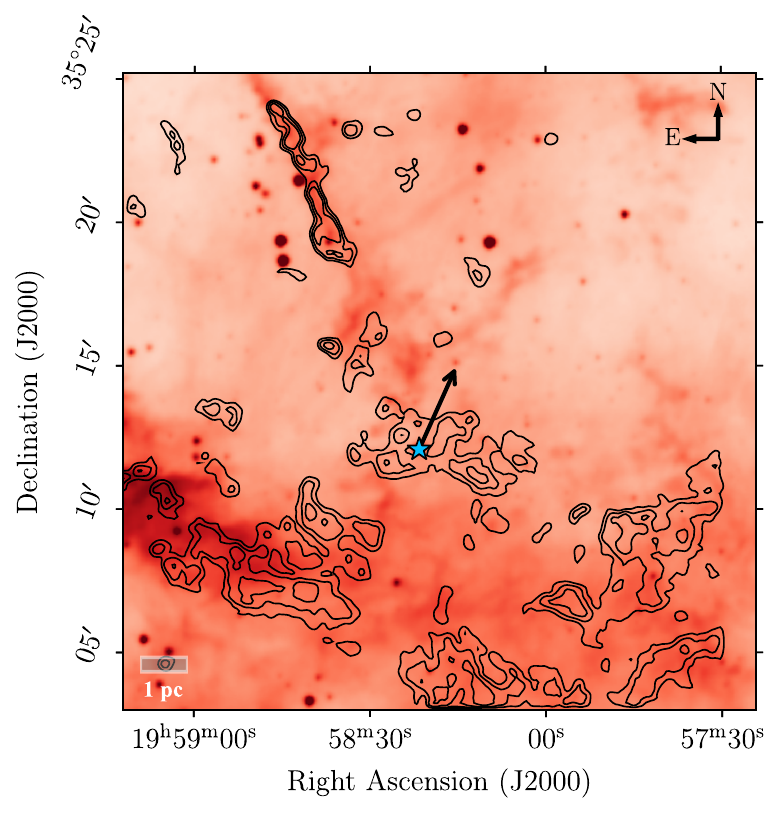}\hfill
    \caption{ WISE observations of the Cyg X--1 region at 3.4 $\mu$m  (W1 filter; \textit{left}) and 12 $\mu$m (W3 filter; \textit{right)}. The position of Cyg X--1 is marked by a blue star and the approaching jet direction is marked with a black arrow. We overlay contours that trace the $^{12}$CO observation with levels $[4.8, 9.6, 16.3]$ K km s$^{-1}$.}
    \label{fig:IR}
\end{figure*}

We include archival images from the Wide-field Infrared Survey Explorer \citep[WISE;][]{2010AJ....140.1868W}  covering the region observed with the IRAM--30m telescope. In Fig~\ref{fig:IR}, we present observations from filters W1 ($3.4\mu$m, \textit{left}, tracing a component mostly related to stellar emission) and W3 ($12\mu$m, \textit{right}, tracing a component mostly related to dust and cold gas). We note that some of the structures observed in molecular gas are also observable at these wavelengths. Namely, \textit{the handlebar} feature near the bow shock (see Fig.~\ref{fig:momentmaps}, \textit{left}) shows a notable abundance of dust as seen in filter W3 (Fig.~\ref{fig:IR}, \textit{right}), and its densest regions correspond to the locations of bright stars in filter W1 (Fig.~\ref{fig:IR}, \textit{left}). This evidence taken together with our spectral analysis (which shows progressively red-shifted peaks from south to north, inconsistent with the action from an approaching jet) likely indicate that this feature could be related to a star forming region unrelated to Cyg X--1.

\section{Column density map} \label{app:columndens}
\begin{figure}[h!]
    \centering
    \includegraphics[width=0.6\textwidth]{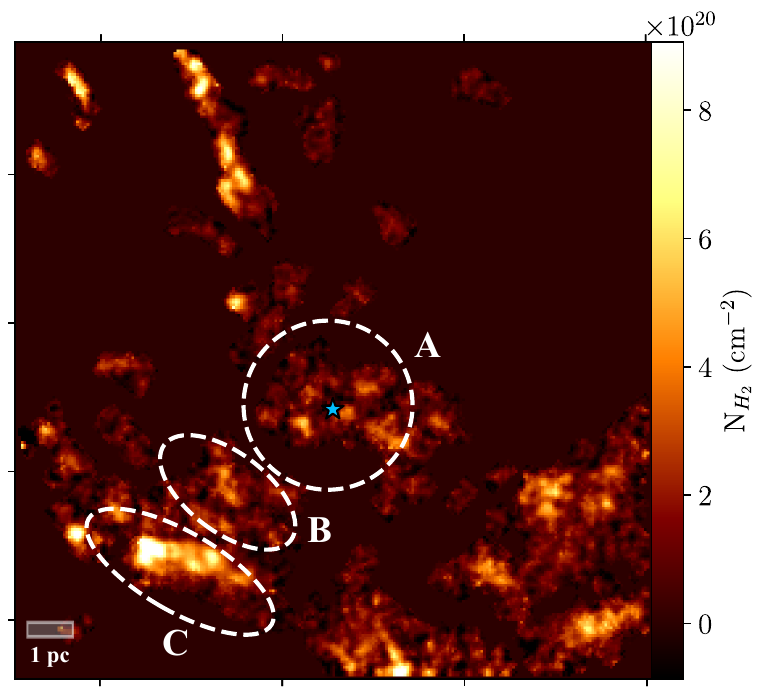}
    \caption{H$_2$ column density maps of the Cyg X--1 region. Local thermal equilibrium was assumed to estimate the optical depth of the region using our $^{12}$CO and $^{13}$CO observations. The position of the central BHXB source is indicated with a blue star marker. We have highlighted regions around \textit{the spider} (A) and the southeastern molecular cloud (B, C) since they are the different components present in Fig.~\ref{fig:PV}.} 
    \label{fig:columndensity}
\end{figure}

We have built a column density map (Fig.~\ref{fig:columndensity}) for $^{13}$CO following the formalism introduced in \cite{wilson2009tools} and previously applied in \cite{tetarenko2018mapping}. We have transformed the $^{13}$CO column density into H$_2$ using the linear ratio presented in \cite{sofue2020co}. We note that the values obtained for H$_2$ column density are within an order of magnitude of those previously reported for hydrogen column density in X-ray spectral fitting \citep{sugimoto2017orbital}. While those values were extracted right on-source and refer to single hydrogen atoms rather than molecular hydrogen, we find similar values in the extended regions covered in this work.

Assuming a distance of 2.22 kpc \citep{miller2021cygnus}, we infer the masses of these three regions with separate velocity peaks:
\begin{itemize}
    \item Region A ({\it the spider}) has a mass of $M_{A}\sim10~\text{M}_\odot$ and has a median velocity of $v_{\rm LSR}\sim7.5~\text{km s}^{-1}$.
    \item Region B has a mass of $M_{B}\sim 7~\text{M}_\odot$, with a peak velocity of $v_{\rm LSR}\sim9~\text{km s}^{-1}$.
    \item  Region C has a mass of $M_{C}\sim 26~\text{M}_\odot$, with a peak velocity of $v_{\rm LSR}\sim14~\text{km s}^{-1}$.
\end{itemize}

\section{Circular path P-V diagrams around other regions} 
\begin{figure*}[h!]
    \centering
    \includegraphics[width=\textwidth]{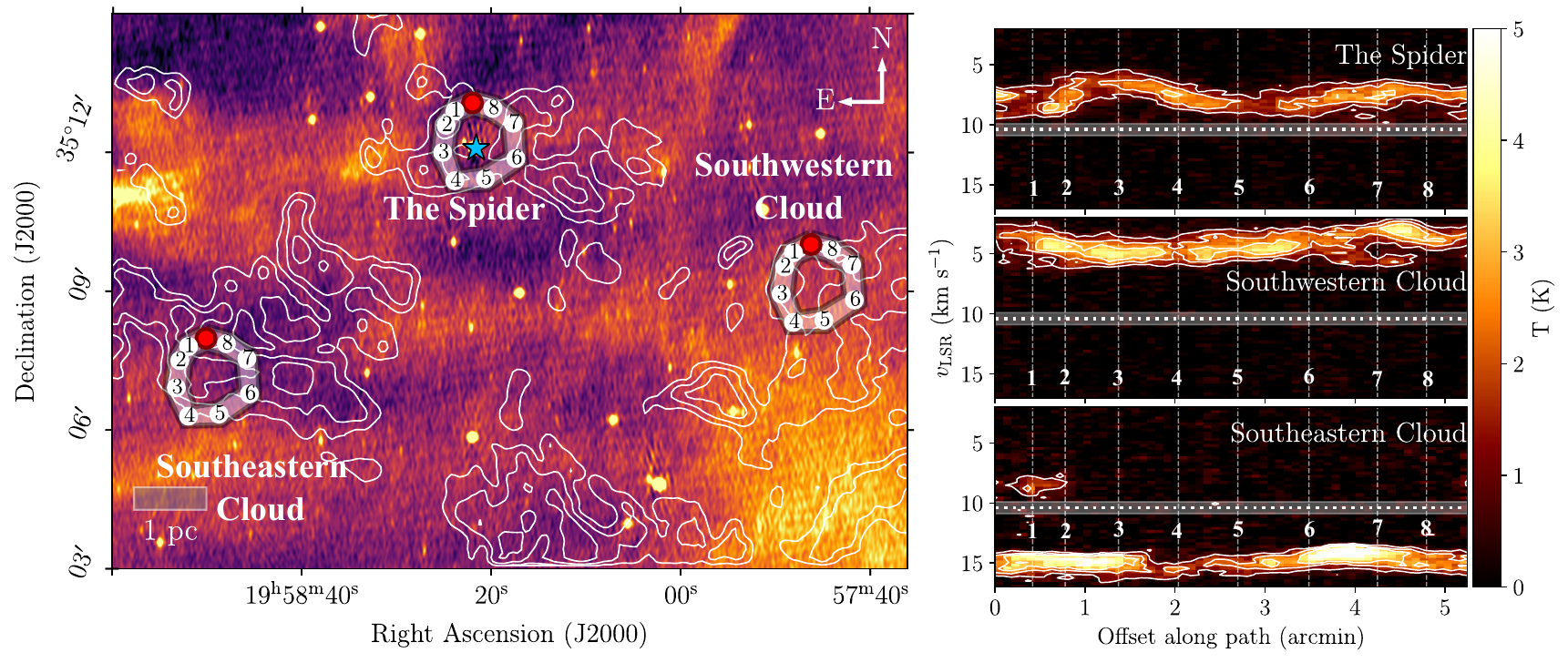}
    \caption{Kinematic analysis of $^{12}$CO($J=2-1$) emission in the Cyg X--1 field. \textit{Left:} Radio continuum (MeerKAT) reference image with the IRAM--30m $^{12}$CO contours overlayed (contour levels [4.8, 9.6, 16.3] K km s$^{-1}$). The BHXB position is marked with a blue star. The $30''$-wide P-V extraction paths are taken along identical closed trajectories around {\it the spider} and the brightest features of the southeastern and southwestern cloud. All paths start from the red marker and proceed counterclockwise, following the numeration. \textit{Right:} $^{12}$CO P-V diagrams along the paths indicated in the \textit{left} panel (contours at [1.0, 2.0, 3.0] K). The horizontal dotted line/gray shading represents the $v_{\text{LSR}}$ of Cyg X--1 with its uncertainty range. The position of the reference points around the highlighted paths are indicated with vertical dashed lines. \textit{The spider}'s pattern shows velocity gradients that are at least $\sim2$ times larger than those found at the comparison regions, and displays two well defined peaks within a circular path.}
    \label{fig:PVcirc}
\end{figure*}

{\label{sec:circ}}
To evaluate the uniqueness of {\it the spider}'s velocity pattern (Fig.~\ref{fig:PV}), we took the same circular path and moved it to the brightest regions in the southeastern and southwestern cloud features (Fig.~\ref{fig:PVcirc}). In comparison to the other two regions probed, {\it the spider} shows a prominent undulation pattern, with clear velocity gradients (of up to $\sim 3~\text{km s}^{-1}~\text{arcmin}^{-1}$) and multiple peaks. We argue that the uniqueness of this velocity pattern is suggestive of an outflow-based interaction (see \S\ref{sec:nonassoc}).

\end{document}